\documentclass[10pt,twocolumn]{revtex4-2} 
\usepackage[dvips]{color}
\usepackage{epsfig}
\usepackage{amssymb}
\usepackage{latexsym}
\usepackage{graphicx}
\usepackage[utf8]{inputenc}
\begin{document}

\title{Lattice simulation of $(2+1)D$ phonetic solitons and the Renormalization group}

\author
{ Sadataka Furui}
\affiliation{(Formerly) Graduate School of Science and Engineering, Teikyo University, Utsunomiya, 320 Japan }
 \email{furui@umb.teikyo-u.ac.jp      }
\date{\today }

\begin{abstract}
The outline of lattice simulations of $(2+1)D$ soliton-propagations in the background of  Weyl spinors is presented. Clifford algebra is applied on Luescher's domain decomposition method. The Clifford algebra shows that there are loop parts and interpolating surface parts in the Wilson's lattice action.

We adopt the Migdal-Kadanoff prescription and the fixed point action in momentum space of Benfatto and Gallavotti, and shows a road map for simulating phonetic solitons in materials.

Detections of topological anomalies (APS index) in nondestructive testing are discussed.
\end{abstract}
\maketitle
\section{ Introduction}
\label{intro}
Propagation of a soliton in backgrounds of fermions is an interesting problem. In nondestructive testing (NDT), 
convolutions of a ultrasonic wave and its time reversed (TR) wave were measured to find anomalies in materials\cite{FDS20,DSP20}.

In the study of many fermion system, Luttinger \cite{Luttinger63} constructed a model whose exact solutions was derived.  Wilson and Fischer \cite{WF72} studied critical exponents of Ising-like $XY$ model, using the fixed point(FP) method.

Bell and Wilson \cite{BW74} studied nonlinear renormalzation group in 2 and 4 fermion interacting Hamiltonian system in momentum space.  Phase transitions in gauge and spin-lattice systems were studied by Migdal \cite{Migdal75a,Migdal75b}. Kadanoff\cite{Kadanoff77} studied lattice renormalization group and the FP phenomenology. Creutz \cite{Creutz80} reviewed the beta function of the Migdal-Kadanoff prescription, and Wilson's lattice simulations \cite{Wilson71a,Wilson71b,Wilson74}.

In order to study Dirac fermion systems with Fermi momentum $p_F=(2\pi/L)(n_F+\frac{1}{2})$ where $[0,L]$ is the domain of a fermion, Benfatto et al. \cite{Gallavotti85,BG90,BGM92,BGPS94,BG95} adopted the renormalization group method. They studied beta function of fermions and bosons in the renormalization group theory.

 Gruzberg et al. \cite{GRV05} showed that in the Altland-Zirnbauer (AZ) symmetry classes $DIII$ \cite{AZ97}, there is a TR symmetry unbroken but spin-rotation (SR) broken phase exists.
Ryu et al.\cite{RML12} studied gravitational and electromagnetic thermal responses using the AZ symmetry classes. In order to analyze whether there is time reversal symmetric spin rotation broken phase exist in bounded fermion background, and whether topological anomaly in fermionic materials can be detected is our problem.
 
Atiyah and Singer defined an index \cite{Hirzebruch78} as the difference of number of zero modes with positive chirality $n+$ and that with negative chirality $n-$, for locally elliptic differential operator $D$ in a compact Riemann manifold. 

The Atiyah-Patodi-Singer (APS) index theorem\cite{APS75}  gives a relation between the number of zero modes in a bounded manifold. Atiyah, Bott and Patodi \cite{ABP73} considered a heat expansions of the index in a bounded region and studied the difference of number of zero mode in $t\geq 0$ region and that in $t< 0$ region. $t\geq 0$ region that in $t<0$ region on Riemann manifolds with boundaries. 
 
I was interested in the APS index as I analyzed the $(2+1)D$ convolution of a ultrasonic non-linear wave propagating in a plane and its time reversed (TR) wave, which is used in non destructive testings (NDT) by Dos Santos' group\cite{DSP20}. 

Nonlinear partial differential equation of Khokhlov-Zaboltskaya (KhZa) has a soliton solution, whose evolution is defined by real variables, and adopted as a model of phonon propagations. 

The ratio of the length at which a discontinuity is formed $l_s=c_0^2/\epsilon\omega u_0$, diffraction length $l_d=\omega a^2/2c_0$ define variables $z=x/l_s$, $N=l_s/l_d$. 

A solution $V=u/u_0$, where $u_0$ is a characteristic amplitude, can be expressed by a series of Bessel functions,
\begin{eqnarray}
V&=&\sum_{n=1}^\infty\frac{2}{n z}J_n(\frac{n z}{\sqrt{ 1+N^2 z^2}}exp(-\frac{R^2}{1+N^2 z^2}))\nonumber\\
&&\times \sin n[\theta+arctan(Nz)-R^2\frac{Nz}{1+{N^2z^2}}]\nonumber
\end{eqnarray}
\begin{figure}
\begin{center}
\includegraphics[width=7cm,angle=0,clip]{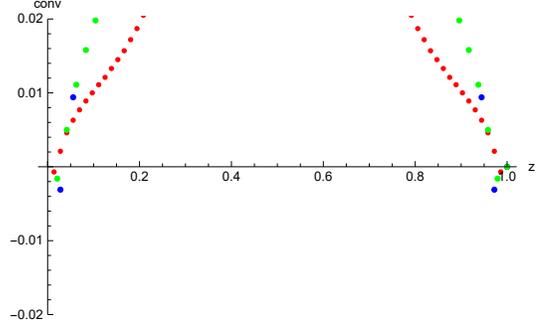}
\end{center}
\caption{The contour of the convolution at $N=0.5$ near boundaries of $z=0$ and $z=1$. Three $\Delta z$ dependence are compared. $\Delta z=1/36$(blue), $1/48$ (green) and $1/72$ (red). }
\label{Convolution}
\end{figure}

A $2D$ convolution as a function of $z$ at $N=0.5$, there are two negative points near the boundary of the Liouville region as shown in Fig.\ref{Convolution}. 
The logaithm of negative convolution has the positive imaginary part. It means that $t\geq 0$ region is stable and $t< 0$ region is unstable, and $n_+-n_-=2$.

In order to preserve chiral symmetry on lattices, Bell and Wilson\cite{BW74} proposed the Fixed Point(FP) action. Luescher and Weisz\cite{LW85} defined improved lattice gauge action following the program of Symanzik. Its action consists of a trial action ${\mathcal A}({\bf S})$ of coarse lattices and  correction terms which are defined recursively. Chiral fermions were studied in \cite{GW81,Luescher98a,Luescher98b,Luescher03,NN93,DGHHN95,HN97,Neuberger97,Hasenfratz98,Niedermeyer99}

The method was applied in two dimensional $O(3) \sigma$ model by Blatter et al.\cite{BBHN96}. The partition function on a lattice is given in a form

\[
Z=\int D{\bf S} e^{-\beta{\mathcal A}({\bf S})},
\]
where the action in the continuum is 
\[
{\mathcal A}_{cont}({\bf S})=\frac{1}{2}\int d^2 x \partial_\mu{\bf S}(x)\cdot\partial_\mu{\bf S}(x).
\]

The structure of this article is as follows.  In the Section II, we explain FP lattice actions in $(2+1)D$.   An explicit calculation of eigenvalues of Wilson loop of length less than or equal to $8a$ is given.
In the Section III. recursive renormalization group analysis of $(2+1)D$ lattices is explained.
In the Section IV, application of Luescher's action of unitary groups  to the symplectic group $C\ell_{1,3}^+$ is explained.   Application of the renormalization group to the scaling problem in propagation of phonetic solitons are discussed in Sction V.  In view of success of instanton picture in spin field theory, relations between the topological charge and instantons and perspective is given in Section VI. 

\section{Fixed point lattice actions in $(2+1)D$}

\subsection{Paths on one 2$D$ plane expanded by $e_1$ and $e_2$}
DeGrand et al. \cite{DGHHN95} adopted FP action in chirally symmetric lattice QCD simulations in $4D$. space-time using $SU(3)$ symmetric Lie groups.
They decomposed the FP action $S^{FP}(V)$ as
\[
S^{FP}(V)=min_{U}(S_0(U)+T(U,V))= S^{FP}(U_0)+T(U_0,V)
\]
where $U_0$ is the fine configuration.  $S_0(U)$ is a selected sets of configutations and $T(U,V)$ is obtained from an average of coarse configurations which is called ${\bf R}$ and $V$,  The series of a configuration $U$ to $U_0$ is chosen by reducing the lattice spacing by a factor of 2.

In this process we replace actions on quarks to actions of Weyl spinors represented by quaternions, and necessarily the FP actions changes to eigenvalues of configurations described by quaternions.

\begin{figure}[h]
\begin{minipage}[b]{0.47\linewidth}
\begin{center}
\includegraphics[width=3.5cm,angle=0,clip]{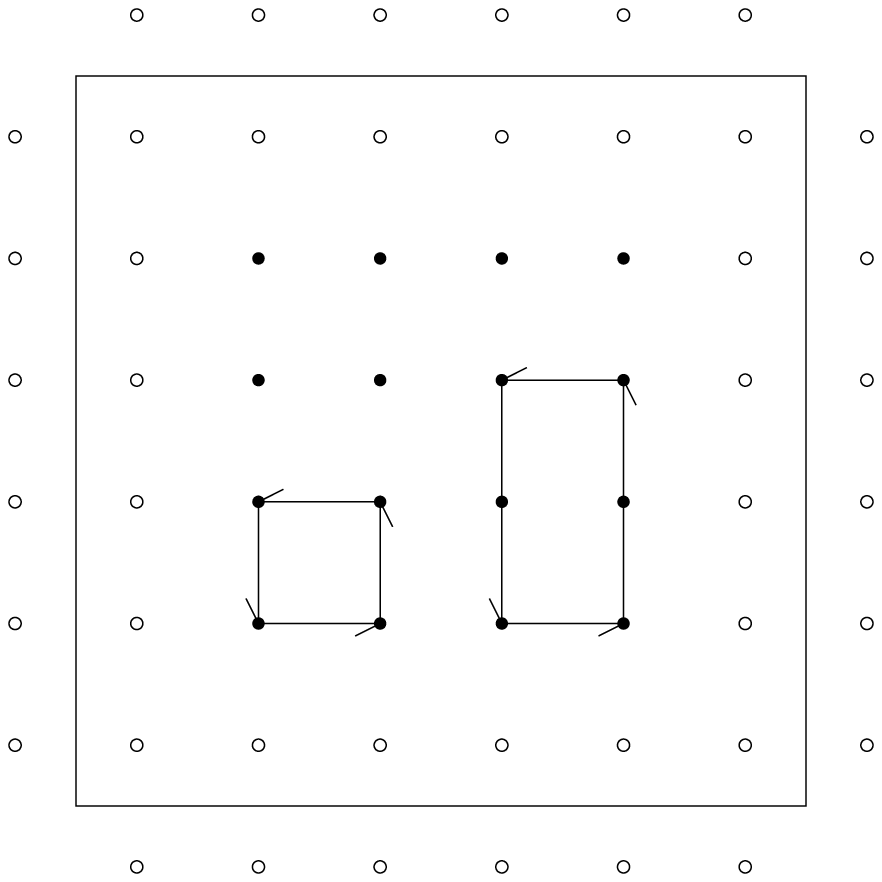} 
\end{center}
\end{minipage}
\hfill
\begin{minipage}[b]{0.47\linewidth}
\begin{center}
\includegraphics[width=3.5cm,angle=0,clip]{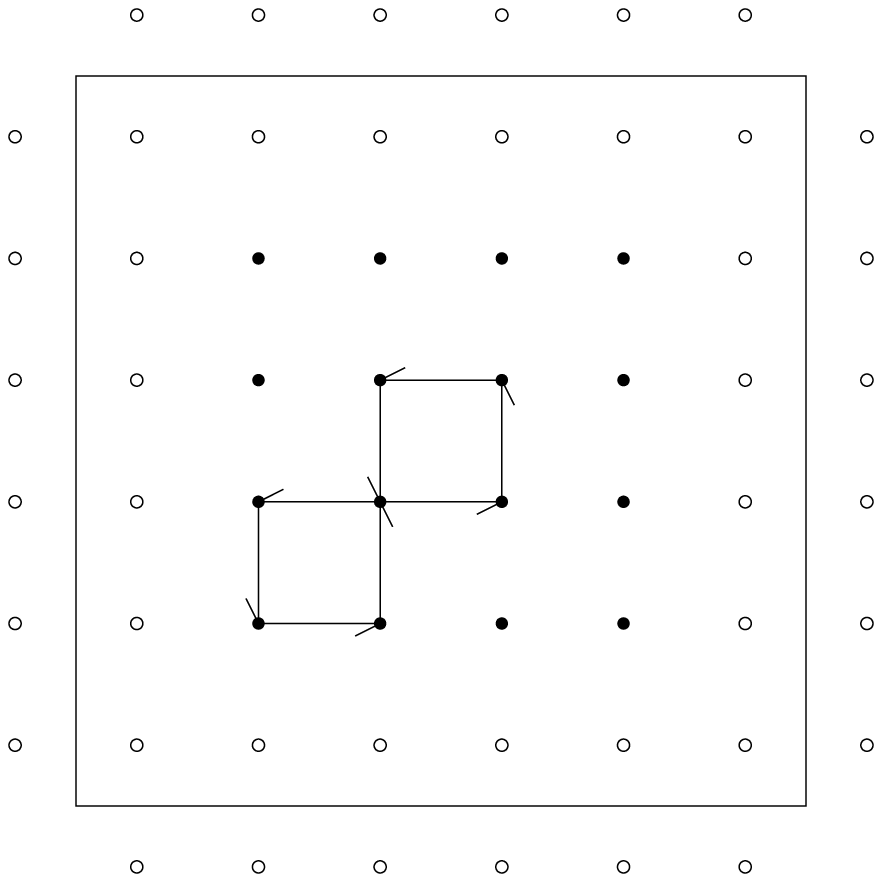} 
\end{center}
\end{minipage}
\caption{ Loop1, Loop2 (left) and Loop5 (right). (Loop28 is a double path of Loop1.)}
\label{Loop125}
\end{figure}

\begin{figure}
\begin{minipage}[b]{0.47\linewidth}
\begin{center}
\includegraphics[width=3.5cm,angle=0,clip]{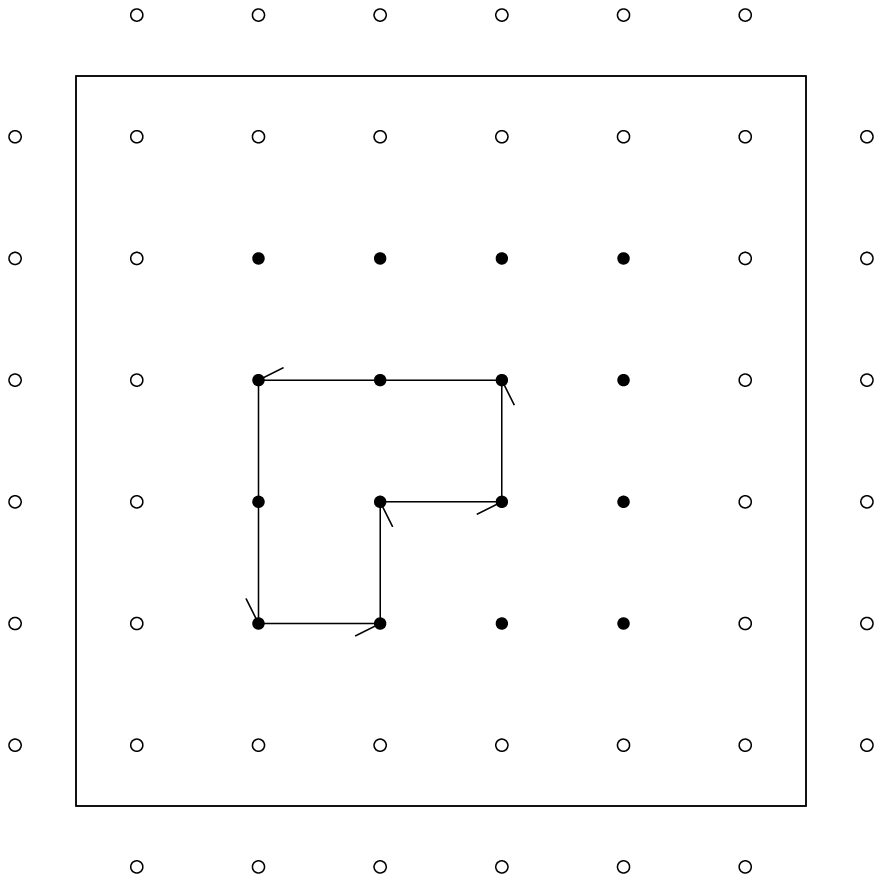} 
\end{center}
\end{minipage}
\hfill
\begin{minipage}[b]{0.47\linewidth}
\begin{center}
\includegraphics[width=3.5cm,angle=0,clip]{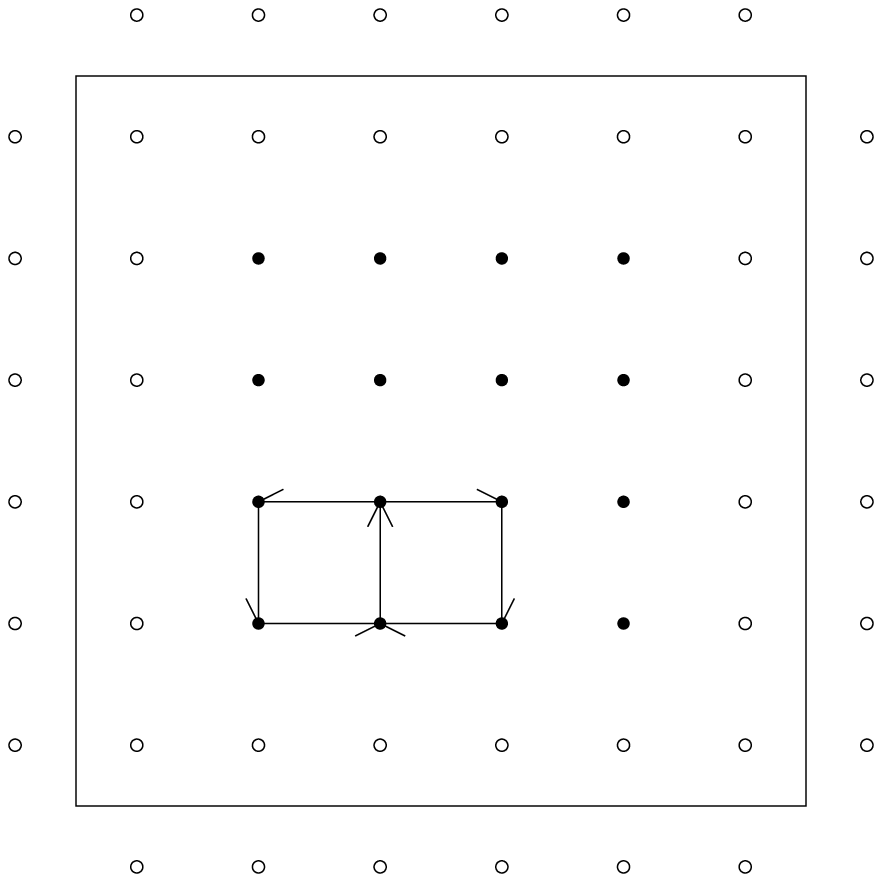} 
\end{center}
\end{minipage}
\caption{ Loop6 (left) and Loop11 (right).}
\label{Loop611}
\end{figure}

\begin{figure}
\begin{minipage}[b]{0.47\linewidth}
\begin{center}
\includegraphics[width=3.5cm,angle=0,clip]{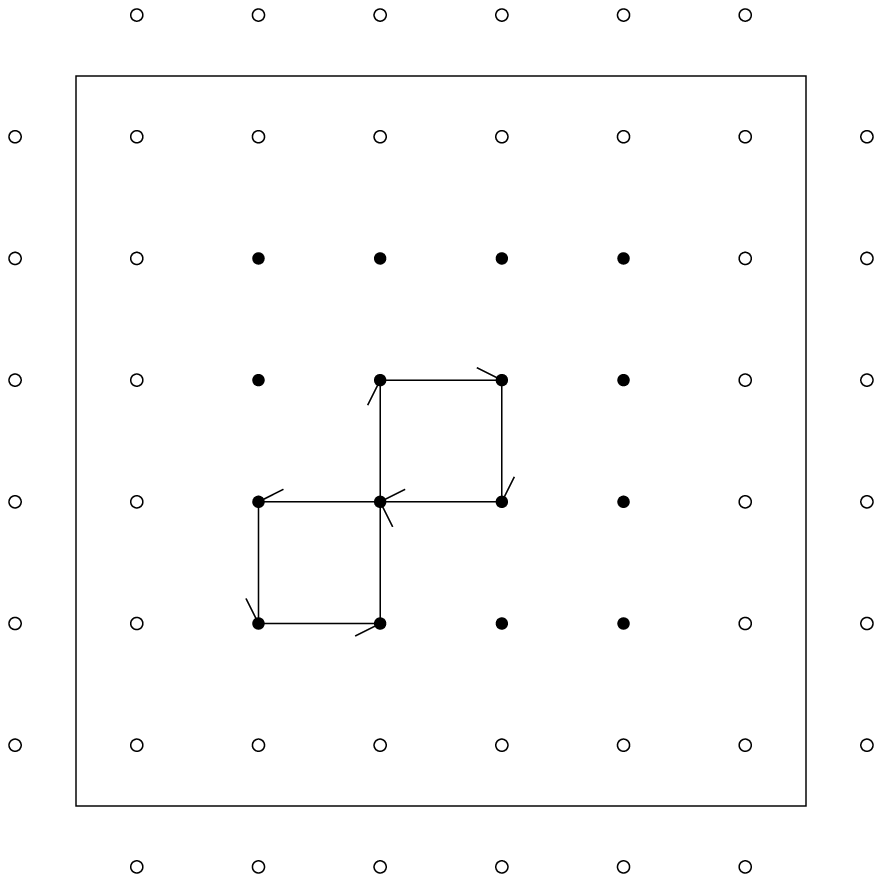} 
\end{center}
\end{minipage}
\hfill
\begin{minipage}[b]{0.47\linewidth}
\begin{center}
\includegraphics[width=3.5cm,angle=0,clip]{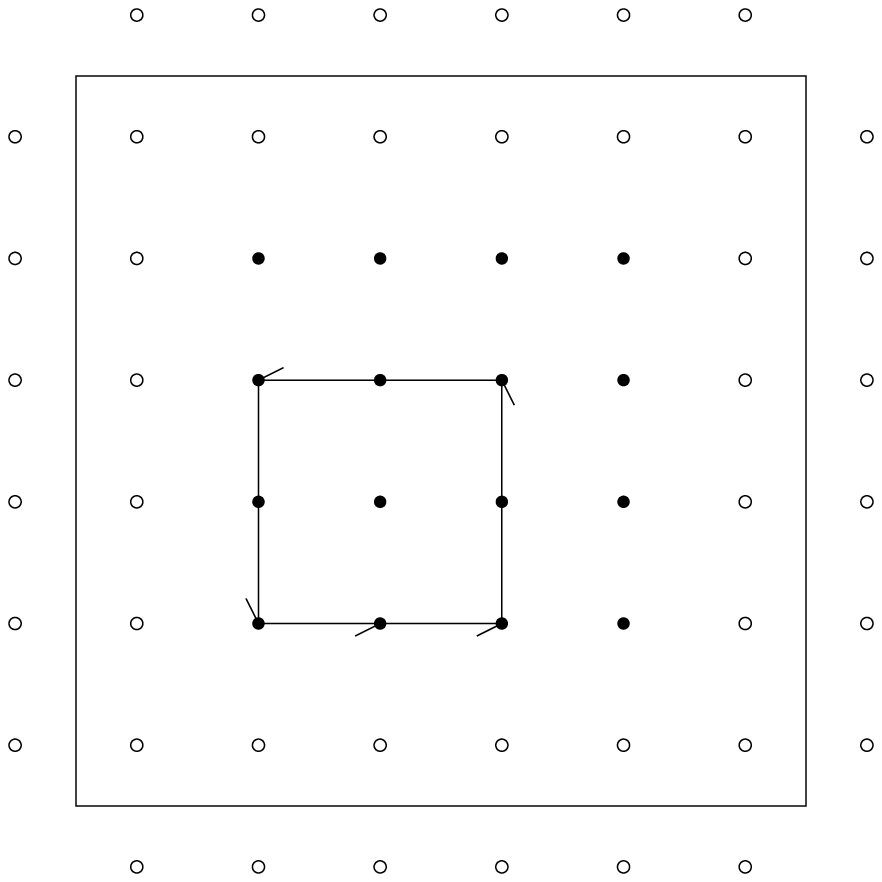} 
\end{center}
\end{minipage}
\caption{ Loop12 (left) and Loop18 (right). ( Path orders of Loop5 and Loop12 are different.) } 
\label{Loop1218}
\end{figure}

The Wilson loops that DeGrand et al. considered relevant to our $(2+1)D$ actions on a 2$D$ plane are shown in Fig.\ref{Loop125}, Fig.\ref{Loop611} and Fig.\ref{Loop1218}.


The Wilson loop plaquette calculation using the Clifford algebra can be performed as follows.

 We define pure quaternions ${\bf S}\in {\bf S}^3$  mapped from $u=u_1+\sqrt{-1}u_2\in{\bf C}$ by
\begin{eqnarray}
&&{ S}^i=\frac{2 u_i}{1+|u|^2}\quad( i=1,2),\quad { S}^0=\frac{1-|u|^2}{1+|u|^2}.\nonumber\\
&&{\bf S}=S^1 e_1+S^2 e_2+S^0 e_1\wedge e_2.\nonumber
\end{eqnarray}
 Here we use $e_1=\left(\begin{array}{cc}
       0& 1\\
       1&0\end{array}\right),\quad e_2=\left(\begin{array}{cc}
                     0 & -\sqrt{-1}\\
                     \sqrt{-1}& 0\end{array}\right)$.

For a $q\in{\bf H}$, taking a conjugation in ${\bf C}$ corresponds to taking a reversion, and we express it by $\tilde q$.

A mapping from a coordinate $x$ in the space ${\bf R}^{1,3 }$ is calculated by using
 $X=\left(\begin{array}{cc}
                                                        x& x x^-\\
                                                        I & x^-\end{array}\right)$,
  where $I=e_0=\left(\begin{array}{cc}
           1&0\\
           0&1\end{array}\right)$ and the hyperplane reflection $x\to -x^-$ is represented by a multiplication of 
   $\left(\begin{array}{cc}
           0&1\\        
          -1&0\end{array}\right)$. 

$e_0^-=-e_0$, $e_1^-=\left(\begin{array}{cc}
 1&0\\
0& -1\end{array}\right)$ and $e_2^-=\left(\begin{array}{cc}
\sqrt{-1}&0\\
0&\sqrt{-1}\end{array}\right)$.

A translation operator ${\mathcal T}=\left(\begin{array}{cc}
I&c\\
0&I\end{array}\right)$, shifts the coordinate $x$ by $c$ \cite{Porteous95}.  The mapping $X \to Y$ is realized by
\[
{\mathcal T}.X.{\mathcal T}^{-1}=\left(\begin{array}{cc}
                     x+c&(x+c)(x^- - c)\\
                     I&x^- -c\end{array}\right)=\left(\begin{array}{cc}
                     y& y y^-\\
                    I& y^-\end{array}\right).
                    \]

We define ${\mathcal T}_{(m,n)}=\left(\begin{array}{cc}
                                      I&(m e_1+n e_2)a\\
                                      0 & I\end{array}\right),$ where $a$ is the lattice constant, and  ${\mathcal T}^{-1}_{(m,n)}={\mathcal T}_{(-m,-n)}$.     
        
 \[
{\mathcal T}_{(m,n)}\left(\begin{array}{cc}
              x(u_1,u_2)& x x^-(u_1,u_2)\\
              I& x^-(u_1,u_2)\end{array}\right){\mathcal T}_{(-m,-n)}
 \]
 agrees with
 \[
\quad\quad\left(\begin{array}{cc}
              x(u_1',u_2')& x x^-(u_1',u_2')\\
              I& x^-(u_1',u_2')\end{array}\right)
\]
where $x(u_1',u_2')$ are $x(u_1+m a,u_2+ n a )$, up to linear order of $a$.                 

Using a mapping of ${\bf H}$ to ${\bf C}$, $X$ can be expressed as
\begin{eqnarray}
&&X(u_1,u_2)=\nonumber\\
&&\left(\begin{array}{cccc}
\sqrt{ -1} u_1&u_2& \sqrt{-1}u_1^2-u_1 u_2& -u_1 u_2+\sqrt{-1}u_2^2\\
-u_2& -\sqrt{-1}u_1& -u_1u_2-\sqrt{-1} u_2^2& \sqrt{-1}u_1^2+u_1 u_2\\
1&0& u_1+\sqrt{-1}u_2&0\\
0&1& 0&u_1+\sqrt{-1} u_2\end{array}\right)\nonumber
 \end{eqnarray}

We calculate the product
${\mathcal T}_{(m,n)} X(u_1, u_2){\mathcal T}_{(-m,-n)}$
using Mathematica\cite{Mathematica} and obtain the link matrix that that operates on quaternions sitting at $(u_1,u_2)$ to another quaternion. 

We define $\partial_1 {\bf S}[u_1,u_2]$ as
\[
d{\bf S}_1[u_1,u_2]=({\mathcal T}_{(1,0)} X(u_1, u_2){\mathcal T}_{(-1,-0)}-X(u_1,u_2))/a.,
\] 
and $\partial_2 {\bf S}[u_1,u_2]$ as
\[
d{\bf S}_2[u_1,u_2]=({\mathcal T}_{(0,1)} X(u_1, u_2){\mathcal T}_{(-0,-1)}-X(u_1,u_2))/a.
\]

If the linked quaternion is separated by $a e_1$, the matrix element is denoted as $t1[a,u_1,u_2]_{ij}$ ($1\leq i,j\leq 4$). 
\begin{eqnarray}
&&t1[a, u_1, u_2]_{11}= \sqrt{-1} (a +\frac{2 u_1}{1 + u_1^2 + u_2^2}),\nonumber\\
&&t1[a, u_1, u_2]_{12}= -\frac{\sqrt{-1} (u_1^2 + (\sqrt{-1} + u_2)^2)}{1 + u_1^2 + u_2^2}, \nonumber\\
&&t1[a, u_1, u_2]_{13}= \frac{ 1}{(1 + u_1^2 + u_2^2)^2}\nonumber\\
&&\times (-8 \sqrt{-1} u_1 u_2 +  a (1 + u_1^2 + u_2^2) (u_1^2 + (-\sqrt{-1} + u_2)^2) \nonumber\\
&& +  a (1 + u_1^2 + u_2^2) (2 u_1 + a (1 + u_1^2 + u_2^2))), \nonumber\\
&&t1[a, u_1, u_2]_{14}=\frac{ 1}{(1 + u_1^2 + u_2^2)^2} (u_1^4 + (\sqrt{-1} + u_2)^4\nonumber\\
&&  + 2 a u_1 (1 + u_1^2 + u_2^2) + 2 u_1^2 (1 + 2 \sqrt{-1} u_2 + u_2^2) \nonumber\\ 
 &&  +a (1 + u_1^2 + u_2^2) (u_1^2 + (\sqrt{-1} + u_2)^2)), \nonumber\\
&&t1[a, u_1, u_2]_{21}=-\frac{\sqrt{-1} (u_1^2 + (-\sqrt{-1} + u_2)^2)}{ 1 + u_1^2 + u_2^2},\nonumber\\
&&t1[a, u_1, u_2]_{22}=  -\sqrt{-1} (a + \frac{2 u_1}{ 1 + u_1^2 + u_2^2}), \nonumber\\
&&t1[a, u_1, u_2]_{23}=  -\frac{1}{(1 + u_1^2 + u_2^2)^2}\nonumber\\
&&\times (u_1^4 + (-\sqrt{-1} + u_2)^4 + 2 a u_1 (1 + u_1^2 + u_2^2)  \nonumber\\
&& +2 u_1^2 (1 - 2 \sqrt{-1} u_2 + u_2^2)  \nonumber\\
&& +  a (1 + u_1^2 + u_2^2) (u_1^2 + (-\sqrt{-1}+ u_2)^2)), \nonumber\\
&&t1[a, u_1, u_2]_{24}= \frac{ 1}{(1 + u_1^2 + u_2^2)^2}\nonumber\\
&&\times (8 \sqrt{-1} u_1 u_2 + 
       a (1 + u_1^2 + u_2^2) (u_1^2 + (\sqrt{-1} + u_2)^2)  \nonumber\\
     &&+  a (1 + u_1^2 + u_2^2) (2 u_1 + a (1 + u_1^2 + u_2^2))),\nonumber\\
&&t1[a, u_1, u_2]_{31}=   1,\nonumber\\ 
&&t1[a, u_1, u_2]_{32}=   0,\nonumber\\
&&t1[a, u_1, u_2]_{33}= -\sqrt{-1}( a+\frac{u_1^2 + (-\sqrt{-1} + u_2)^2}{1 + u_1^2 + u_2^2}),\nonumber\\
&&t1[a, u_1, u_2]_{34}= -\frac{2\sqrt{-1} u_1}{ 1 + u_1^2 + u_2^2},\nonumber\\
&&t1[a, u_1, u_2]_{41}=0,\nonumber\\
&&t1[a, u_1, u_2]_{42}= 1,\nonumber\\
&&t1[a, u_1, u_2]_{43}= -\frac{2 \sqrt{-1} u_1}{1 + u_1^2 + u_2^2}, \nonumber\\
&&t1[a, u_1, u_2]_{44}=\sqrt{-1} (a +\frac{ (u_1^2 + (\sqrt{-1} + u_2)^2}{1 + u_1^2 + u_2^2}).\nonumber
\end{eqnarray}

Similarly link to another quaternion separated by $b e_2$ is represented by
\begin{eqnarray}
&&t2[b,u_1,u_2]_{11}=\frac{2 \sqrt{-1} u_1}{1 + u_1^2 + u_2^2}, \nonumber\\
&&t2[b,u_1,u_2]_{12}= b - \sqrt{-1}\frac{u_1^2 + (\sqrt{-1} + u_2)^2}{1 + u_1^2 + u_2^2},\nonumber\\
&&t2[b,u_1,u_2]_{13}= b^2 - \frac{8 \sqrt{-1}u_1 u_2}{(1 + u_1^2 + u_2^2)^2}\nonumber\\
&& -\frac{ \sqrt{-1} b (2 u_1 + u_1^2 + (\sqrt{-1} + u_2)^2)}{1 + u_1^2 + u_2^2}, \nonumber\\
&&t2[b,u_1,u_2]_{14}=\frac{ 1}{(1 + u_1^2 + u_2^2)^2}\nonumber\\
&& \times (u_1^4 + (\sqrt{-1} + u_2)^4 - 2 \sqrt{-1} b u_1 (1 + u_1^2 + u_2^2) \nonumber\\
&& +2 u_1^2 (1 + 2 \sqrt{-1} u_2 + u_2^2)  \nonumber\\
&&+\sqrt{-1} b (1 + u_1^2 + u_2^2) (u_1^2 + (\sqrt{-1} + u_2)^2)), \nonumber\\
&&t2[b,u_1,u_2]_{21}= -b -\frac{ \sqrt{-1} (u_1^2 + (-\sqrt{-1} + u_2)^2)}{1 + u_1^2 + u_2^2}, \nonumber\\
&&t2[b,u_1,u_2]_{22}=-\frac{2\sqrt{-1} u_1}{ 1 + u_1^2 +   u_2^2}, \nonumber\\
&&t2[b,u_1,u_2]_{23}= -\frac{1}{(1 + u_1^2 + u_2^2)^2}\nonumber\\
&&\times (u_1^4 + (-\sqrt{-1} + u_2)^4 +  2 \sqrt{-1} b u_1 (1 + u_1^2 + u_2^2) \nonumber\\
&&+ 2 u_1^2 (1 - 2 \sqrt{-1} u_2 + u_2^2) \nonumber\\
&&-  \sqrt{-1} b (1 + u_1^2 + u_2^2) (u_1^2 + (-\sqrt{-1} + u_2)^2)),\nonumber\\ 
&&t2[b,u_1,u_2]_{24}= \frac{ 1}{(1 + u_1^2 + u_2^2)^2}\nonumber\\
&&\times (8\sqrt{-1} u_1 u_2 + 2 \sqrt{-1} b u_1 (1 + u_1^2 + u_2^2)  \nonumber\\
 &&+ b (1 + u_1^2 + u_2^2) (b (1 + u_1^2 + u_2^2) \nonumber\\
 &&+ \sqrt{-1} (u_1^2 + (-\sqrt{-1} + u_2)^2))),\nonumber\\       
&&t2[b,u_1,u_2]_{31}= 1,\nonumber\\ 
&&t2[b,u_1,u_2]_{32}= 0,\nonumber\\
&&t2[b,u_1,u_2]_{33}=-\frac{\sqrt{-1} (u_1^2 + (-\sqrt{-1} + u_2)^2)}{1 + u_1^2 + u_2^2},\nonumber\\
&&t2[b,u_1,u_2]_{34}= -b - \frac{2\sqrt{-1} u_1}{ 1 + u_1^2 + u_2^2}, \nonumber\\
&&t2[b,u_1,u_2]_{41}=0,\nonumber\\
&&t2[b,u_1,u_2]_{42}=1,\nonumber\\
&&t2[b,u_1,u_2]_{43}= b -\frac{2 \sqrt{-1} u_1}{1 + u_1^2 + u_2^2},\nonumber\\
&&t2[b,u_1,u_2]_{44}= \frac{ \sqrt{-1}(u_1^2 + (\sqrt{-1} + u_2)^2)}{1 + u_1^2 + u_2^2}.\nonumber
\end{eqnarray}
    
\begin{figure}[htb]
\begin{center}
\includegraphics[width=7cm,angle=0,clip]{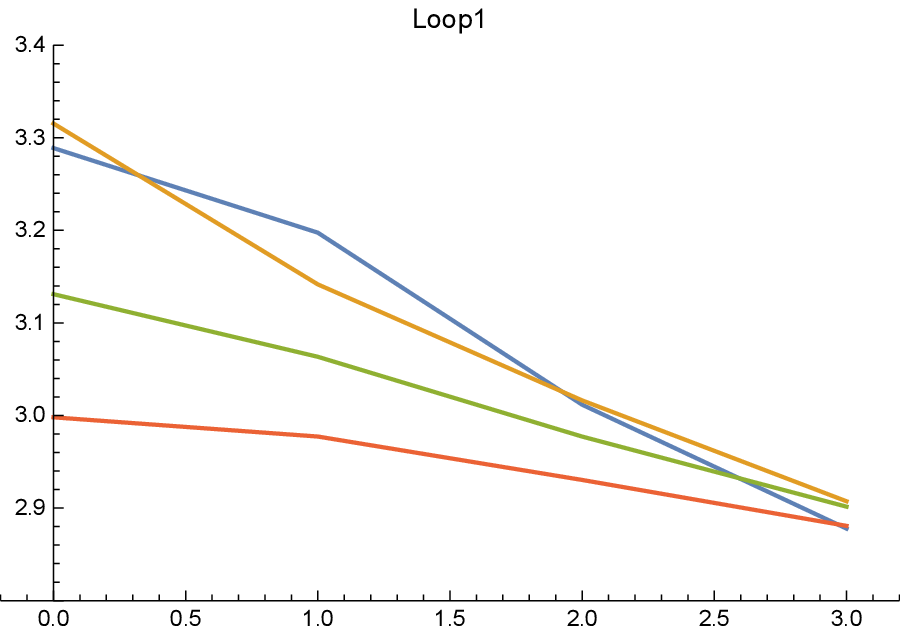} 
\end{center}
\caption{ The absolute value of eigenvalues in Loop1 for a fixed $u_1$ as a function of $u_2$. ($\Delta u_i=1$ $(i=1,2)$) }
\label{L1}
\end{figure}

For the Loop1, we multiply matrices
\begin{eqnarray}
&&L1[u_1,u_2]=t2[-\frac{1}{4},u_1,u_2+\frac{1}{4}]\times t1[-\frac{1}{4}, u_1 + \frac{1}{4}, u_2 + \frac{1}{4}]\nonumber\\
 &&\times t2[\frac{1}{4}, u_1 + \frac{1}{4}, u_2]\times t1[\frac{1}{4}, u_1, u_2].\nonumber
\end{eqnarray}
  The absolute value of the eigenvalues of $d{\bf S}[0, i]$ (Blue),$d{\bf S}[1, i]$ (Orange),$d{\bf S}[2,i]$ (Green), $d{\bf S}[3,i]$ (Red) for $i=0,1,2,3$ are plotted in Fig,\ref{L1}. The similar figures are shown in the following loops. 

For the Loop 2
\begin{eqnarray}
&&L2[u_1,u_2]=t2[-\frac{1}{2}, u_1, u_2 +\frac{1}{2}]\times t1[-\frac{1}{4}, u_1 + \frac{1}{4}, u_2 +\frac{1}{2}]\nonumber\\
&&\times t2[\frac{1}{2}, u_1 + \frac{1}{4},  u_2]\times t1[\frac{1}{4}, u_1, u_2].\nonumber
\end{eqnarray}
\begin{figure}[htb]
\begin{center}
\includegraphics[width=7cm,angle=0,clip]{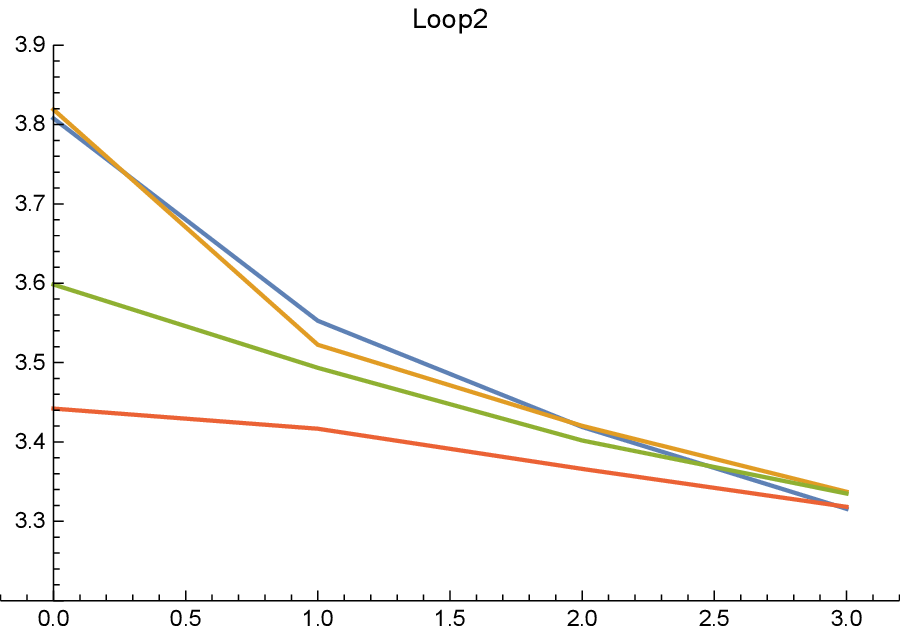} 
\end{center}
\caption{ The absolute value of eigenvalues in Loop2 for a fixed $u_1$ as a function of $u_2$. ($\Delta u_i=1$ $(i=1,2)$) }
\end{figure}

For the Loop5
\begin{eqnarray}
&&L5[u_1,u_2]=t2[-\frac{1}{4}, u_1, u_2 +\frac{1}{4}]\times t1[-\frac{1}{4}, u_1, u_2 +\frac{1}{4}]\nonumber\\
&&\times t2[-\frac{1}{4}, u_1 +\frac{1}{4},  u_2 + \frac{1}{4}]\times t1[-\frac{1}{4}, u_1 + \frac{1}{4}, u_2 + \frac{1}{2}]\nonumber\\
&&\times t2[\frac{1}{4}, u_1 + \frac{1}{2},  u_2 +\frac{1}{2}]\times t1[\frac{1}{4}, u_1 +\frac{1}{4}, u_2 + \frac{1}{4}]\nonumber\\
&&\times t2[\frac{1}{4}, u_1 + \frac{1}{4}, u_2]\times t1[\frac{1}{4},   u_1, u_2].\nonumber
\end{eqnarray}
\begin{figure}
\begin{center}
\includegraphics[width=7cm,angle=0,clip]{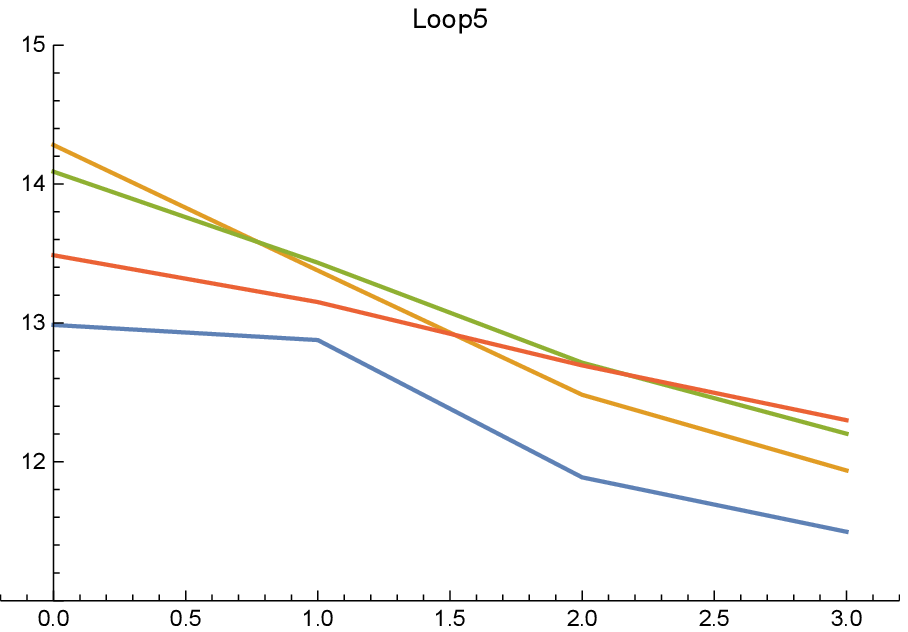} 
\end{center}
\caption{ The absolute value of eigenvalues in Loop5 for a fixed $u_1$ as a function of $u_2$. ($\Delta u_i=1$ $(i=1,2)$) }
\end{figure}

For the Loop 6
\begin{eqnarray}
&&L6[u_1,u_2]=t2[-\frac{1}{2}, u_1, u_2 +\frac{1}{2}]\times t1[-\frac{1}{2}, u_1 + \frac{1}{2}, u_2 + \frac{1}{2}]\nonumber\\
&&\times t2[\frac{1}{4}, u_1 + \frac{1}{2}, u_2 + \frac{1}{4}]\times t1[\frac{1}{4}, u_1 + \frac{1}{4}, u_2 +\frac{1}{4}]\nonumber\\
&&\times t2[\frac{1}{4}, u_1 + \frac{1}{4}, u_2]\times t1[\frac{1}{4}, u_1, u_2].\nonumber
\end{eqnarray}
\begin{figure}[htb]
\begin{center}
\includegraphics[width=7cm,angle=0,clip]{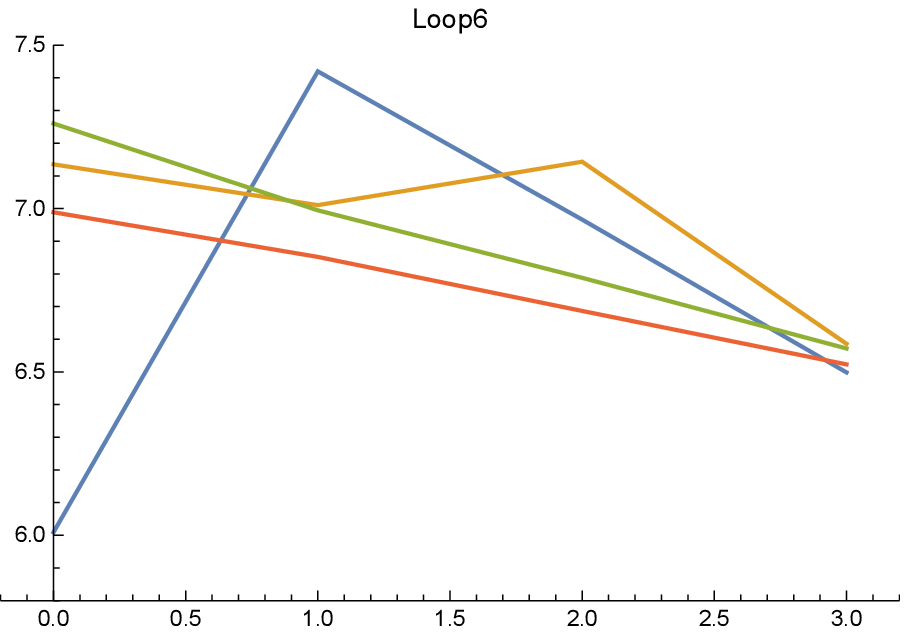} 
\end{center}
\caption{ The absolute value of eigenvalues in Loop6 for a fixed $u_1$ as a function of $u_2$. ($\Delta u_i=1$ $(i=1,2)$) .}
\end{figure}

For the Loop11
\begin{eqnarray}
&&L11[u_1,u_2]=t2[-\frac{1}{4}, u_1, u_2 + \frac{1}{4}]\times t1[-\frac{1}{4}, u_1 +\frac{1}{4}, u_2 +\frac{1}{4}]\nonumber\\
&&\times t2[\frac{1}{4}, u_1 +\frac{1}{4},   u_2]\times t1[-\frac{1}{4}, u_1 + \frac{1}{2}, u_2]\nonumber\\
  &&\times t2[-\frac{1}{4}, u_1 +\frac{1}{2}, u_2 +\frac{1}{4}]\times t1[\frac{1}{4}, u_1 +\frac{1}{4}, u_2 +\frac{1}{4}]\nonumber\\
  &&\times t2[\frac{1}{4}, u_1 + \frac{1}{4}, u_2]\times t1[\frac{1}{4}, u_1, u_2].\nonumber
\end{eqnarray}
\begin{figure}[htb]
\begin{center}
\includegraphics[width=7cm,angle=0,clip]{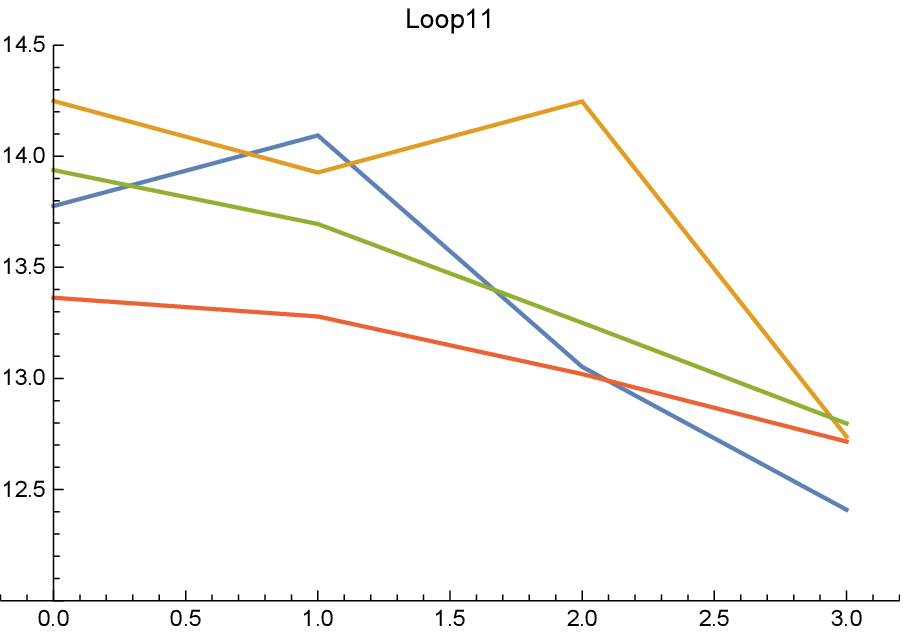} 
\end{center}
\caption{ The absolute value of eigenvalues in Loop11 for a fixed $u_1$ as a function of $u_2$. ($\Delta u_i=1$ $(i=1,2)$).}
\end{figure}

For the Loop12
\begin{eqnarray}
&&L12[u_1,u_2]=t2[-\frac{1}{4}, u_1, u_2 +\frac{ 1}{4}]\times t1[-\frac{1}{2}, u_1 + \frac{1}{2}, u_2 +\frac{1}{4}]\nonumber\\
&&\times t2[-\frac{1}{4}, u_1 +\frac{1}{2}, u_2 +\frac{1}{2}]\times t1[\frac{1}{4}, u_1 + \frac{1}{4}, u_2 +\frac{1}{2}]\nonumber\\
&&\times t2[\frac{1}{2}, u_1 +\frac{1}{4}, u_2]\times t1[\frac{1}{4}, u_1, u_2].\nonumber
\end{eqnarray}
\begin{figure}[htb]
\begin{center}
\includegraphics[width=7cm,angle=0,clip]{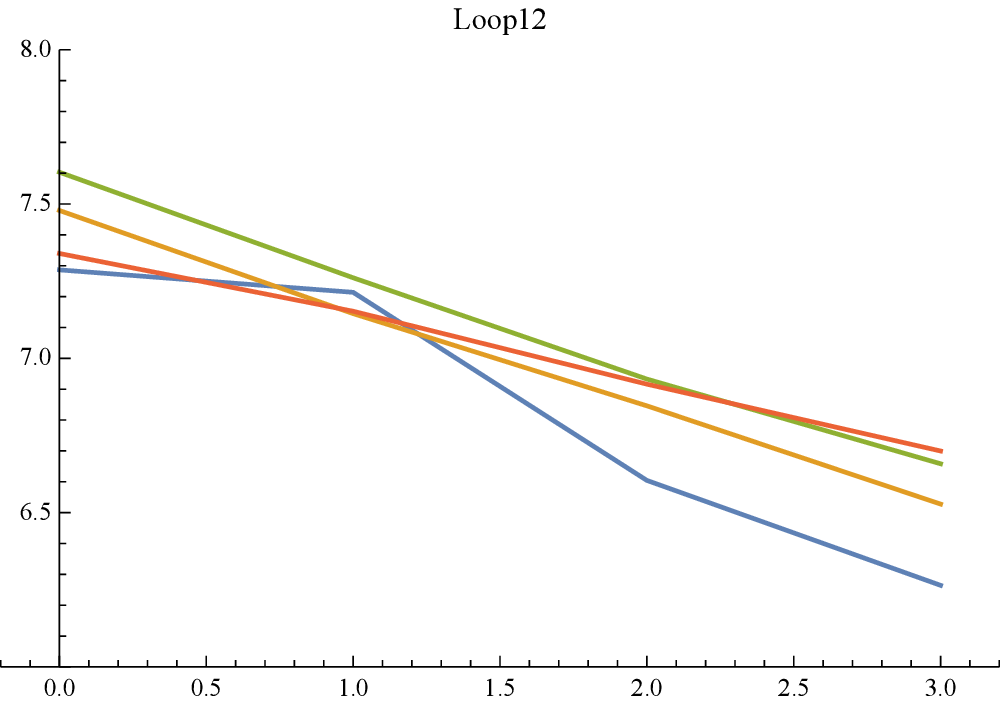} 
\end{center}
\caption{ The absolute value of eigenvalues in Loop12 for a fixed $u_1$ as a function of $u_2$. ($\Delta u_i=1$ $(i=1,2)$) }
\end{figure}

For the Loop18
 \begin{eqnarray}
&&L18[u_1,u_2]= t2[-\frac{1}{2}, u_1, u_2 +\frac{1}{2}]\times t1[-\frac{1}{2}, u_1 + \frac{1}{2}, u_2 + \frac{1}{2}]\nonumber\\
 &&\times t2[\frac{1}{2}, u_1 +\frac{1}{2}, u_2]\times t1[\frac{1}{2}, u_1, u_2].\nonumber
  \end{eqnarray}

\begin{figure}[htb]
\begin{center}
\includegraphics[width=7cm,angle=0,clip]{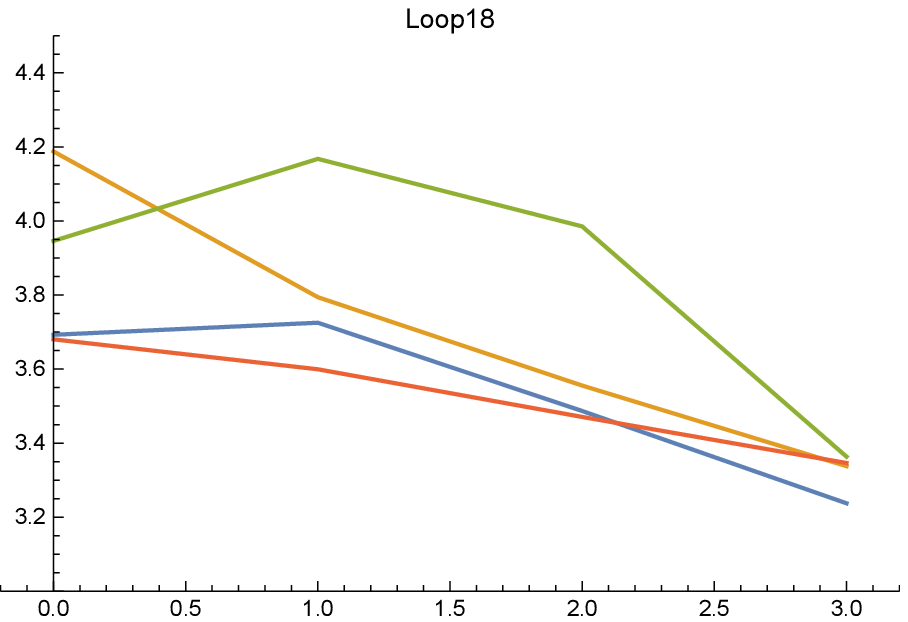} 
\end{center}
\caption{ The absolute value of eigenvalues in Loop18 for a fixed $u_1$ as a function of $u_2$. ($\Delta u_i=1$ $(i=1,2)$) }
\label{L18}
\end{figure}

  For the Loop28
  \begin{eqnarray} 
  &&L28[u_1,u_2]=t2[-\frac{1}{4}, u_1, u_2 +\frac{ 1}{4}]\times t1[-\frac{1}{4}, u_1 +\frac{1}{4}, u_2 + \frac{1}{4}]\nonumber\\
  &&\times t2[\frac{1}{4}, u_1 +\frac{1}{4}, u_2]\times t1[\frac{1}{4}, u_1, u_2]\nonumber\\
  &&\times t2[-\frac{1}{4}, u_1, u_2 + \frac{1}{4}]\times t1[-\frac{1}{4}, u_1 + \frac{1}{4}, u_2 +\frac{1}{4}]\nonumber\\
 &&\times t2[\frac{1}{4}, u_1 +\frac{1}{4}, u_2]\times t1[\frac{1}{4}, u_1, u_2].\nonumber
\end{eqnarray}
\begin{figure}[htb]
\begin{center}
\includegraphics[width=7cm,angle=0,clip]{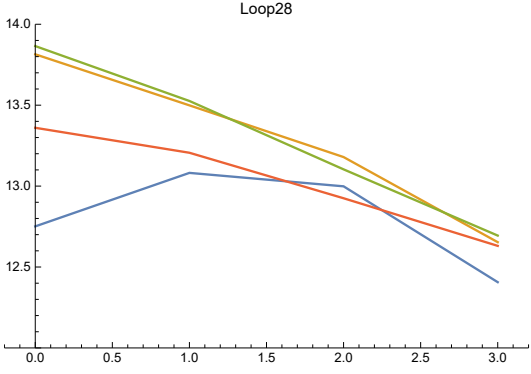} 
\end{center}
\caption{ The absolute value of eigenvalues in Loop28 for a fixed $u_1$ as a function of $u_2$. ($\Delta u_i=1$ $(i=1,2)$)}
\label{L28}
\end{figure}

     The eigen-quatrnions give the fixed point on the manifold.
     
 As an example the $2\times 2$ matrix at left-down corner of the $4\times 4$ matrix of $L1[0,0]$ has the eigenvalue
  \[ 
  \chi(L1[0,0])=1.98353 \pm 2.62338\sqrt{-1}
  \]
   and the eigen-quaternion
  \begin{eqnarray}
 &&{\bf S}(L1[0,0])\nonumber\\
 &=&\left(\begin{array}{cc}
 0.97473 & 0.20534 + 0.087986\sqrt{-1}\\
 -0.20534 + 0.087986 \sqrt{-1}& 0.97473 
 \end{array}\right)\nonumber\\
 &=&0.97473 e_0+0.20534 e_2+0.087986 e_1\wedge e_2.\nonumber
  \end{eqnarray}

Although the path is on a $2D$ plane, the eigenvector has the $e_1\wedge e_2$ component. 

When $u_1$ and $u_2$ are large, convergence of eigenvalues can be seen. When a path intersects to its own path, the convergence at large ${u_1}^2+{ u_2}^2$ is not strong.
  
Along the path $L1[0,0]$,
\[
d{\bf S}(L1[0,0])=dS_1[0,0]+dS_2[\frac{1}{4},0]-dS_1[\frac{1}{4},\frac{1}{4}]-dS_2[0,\frac{1}{4}]
\]
has a non-zero quaternion element in the right-upper corner of the $4\times 4$ matrix.,
\begin{eqnarray}
&&\left(\begin{array}{cc}
 -1.13725 - 0.02614\sqrt{-1}& -0.19608 - 0.91503\sqrt{-1}\\
 0.19608 - 0.91503\sqrt{-1}& -1.13725 + 0.02614\sqrt{-1}\end{array}\right)\nonumber\\
&=&-1.13725e_0-0.02614e_1-0.19608e_2-0.91503 e_1\wedge e_2, \nonumber
\end{eqnarray}
and the other elements are zero.

There are 9 configurations of $d{\bf S}(L1[u_i,u_j])$,  $(0\leq i/4,j/4\leq 1)$.

In the case of Loop18, whose length of the side is twice as that of Loop1,
   \[ 
  \chi(L18[0,0])=2.31556 \pm 2.87632\sqrt{-1}
  \]
and the eigen-quaternion
  \begin{eqnarray}
 &&{\bf S}(L18[0,0])\nonumber\\
 &=&\left(\begin{array}{cc}
 0.25767 + 0.14704 \sqrt{-1} & 0.95498 \\
0.95498 & -0.25767 + 0.14704 \sqrt{-1}
 \end{array}\right)\nonumber\\
 &=&0.14704\sqrt{-1} e_0-0.25267\sqrt{-1} e_1-0.95498\sqrt{-1} e_1\wedge e_2,\nonumber
  \end{eqnarray}
  \[
d{\bf S}(L18[0,0])=dS_1[0,0]+dS_2[\frac{1}{2},0]-dS_1[\frac{1}{2},\frac{1}{2}]-dS_2[0,\frac{1}{2}]
\]
has the right-upper corner
\begin{eqnarray}
&&\left(\begin{array}{cc}
 -2.13333 - 0.13333\sqrt{-1}& -0.53333 - 1.46667\sqrt{-1}\\
 0.53333 - 1.46667\sqrt{-1}& -2.13333 + 0.13333 \sqrt{-1}\end{array}\right)
 \nonumber\\
&=&-2.1333e_0-0.13333 e_1-0.53333e_2-1.46667e_1\wedge e_2.\nonumber
\end{eqnarray}

There are 4 configurations of $d{\bf S}(L18[u_i,u_j])$
$(u_i,u_j)=(0,0),(\frac{1}{2},0)$,$(\frac{1}{4},\frac{1}{4})$,$(0,\frac{1}{4})$.
The absolute value of the eigenvalues of $L18$ are enhanced from that of $L1$.
Elements of eigen-quaternions of $L18$ are multiplied by pure imaginary numbers 
to those of $L1$.
\subsection{Paths on two planes connected by $e_1\wedge e_2$}
The paths containing $e_1\wedge e_2$ components are analyzed similarly to that without containing the terms.  

The link to another quaternion separated by $c e_1\wedge e_2$ is represented by
\begin{eqnarray}
&&t3[c,u_1,u_2]_{11}=\frac{2 \sqrt{-1} u_1}{1 + u_1^2 + u_2^2}, \nonumber\\
&&t3[c,u_1,u_2]_{12}= \sqrt{-1}(c - \frac{u_1^2 + (\sqrt{-1} + u_2)^2}{1 + u_1^2 + u_2^2}),\nonumber\\
&&t3[c,u_1,u_2]_{13}= \frac{1}{(1 + u_1^2 + u_2^2)^2}\nonumber\\
&&\times (-8 \sqrt{-1}  u_1 u_2 +2 c u_1 (1 + u_1^2 + u_2^2)\nonumber\\
&&-c (1 + u_1^2 + u_2^2) (u_1^2 + (\sqrt{-1} + u_2)^2\nonumber\\
&&- c (1 + u_1^2 + u_2^2))), \nonumber\\
&&t3[c,u_1,u_2]_{14}=\frac{ 1}{(1 + u_1^2 + u_2^2)^2}\nonumber\\
&& \times(u_1^4 + (\sqrt{-1} + u_2)^4 + 2 c u_1 (1 + u_1^2 + u_2^2) \nonumber\\
&&+ 2 u_1^2 (1 + 2 \sqrt{-1} u_2 + u_2^2) \nonumber\\
&&- c (1 + u_1^2 + u_2^2) (u_1^2 + (-\sqrt{-1} + u_2)^2)), \nonumber\\
&&t3[c,u_1,u_2]_{21}= \sqrt{-1}(c -\frac{ \sqrt{-1} (u_1^2 + (-\sqrt{-1} + u_2)^2)}{1 + u_1^2 + u_2^2}, \nonumber\\
&&t3[c,u_1,u_2]_{22}=-\frac{2\sqrt{-1} u_1}{ 1 + u_1^2 +   u_2^2}, \nonumber\\
&&t3[c,u_1,u_2]_{23}= -\frac{1}{(1 + u_1^2 + u_2^2)^2}\nonumber\\
&&\times (u_1^4 + (-\sqrt{-1} + u_2)^4 +  2 \sqrt{-1} c u_1 (1 + u_1^2 + u_2^2)\nonumber\\
&& - c (1 + u_1^2 + u_2^2) (u_1^2 + (-\sqrt{-1} + u_2)^2)),\nonumber\\ 
&&t3[c,u_1,u_2]_{24}= \frac{ 1}{(1 + u_1^2 + u_2^2)^2}\nonumber\\
&&\times (8\sqrt{-1} u_1 u_2 + 2 c u_1 (1 + u_1^2 + u_2^2) \nonumber\\
&&-  c (1 + u_1^2 + u_2^2) (u_1^2 + (-\sqrt{-1} + u_2)^2 - c (1 + u_1^2 + u_2^2))) ,\nonumber\\     
&&t3[c,u_1,u_2]_{31}= 1,\nonumber\\ 
&&t3[c,u_1,u_2]_{32}= 0,\nonumber\\
&&t3[c,u_1,u_2]_{33}=-\frac{\sqrt{-1} (u_1^2 + (-\sqrt{-1} + u_2)^2)}{1 + u_1^2 + u_2^2},\nonumber\\
&&t3[c,u_1,u_2]_{34}= -\sqrt{-1}(c + \frac{2 u_1}{ 1 + u_1^2 + u_2^2}), \nonumber\\
&&t3[c,u_1,u_2]_{41}=0,\nonumber\\
&&t3[c,u_1,u_2]_{42}=1,\nonumber\\
&&t3[c,u_1,u_2]_{43}= -\sqrt{-1}(c+\frac{2  u_1}{1 + u_1^2 + u_2^2}),\nonumber\\
&&t3[c,u_1,u_2]_{44}= \frac{ \sqrt{-1}(u_1^2 + (\sqrt{-1} + u_2)^2)}{1 + u_1^2 + u_2^2}.\nonumber
\end{eqnarray}
We define $\partial_3 {\bf S}[u_1,u_2]$ as
\[
\partial_3{\bf S}[u_1,u_2]=({\mathcal T}_{(1,1)} X(u_1, u_2){\mathcal T}_{(-1,-1)}-X(u_1,u_2))/a.
\]

\begin{figure}[h]
\begin{minipage}[b]{0.47\linewidth}
\begin{center}
\includegraphics[width=3.5cm,angle=0,clip]{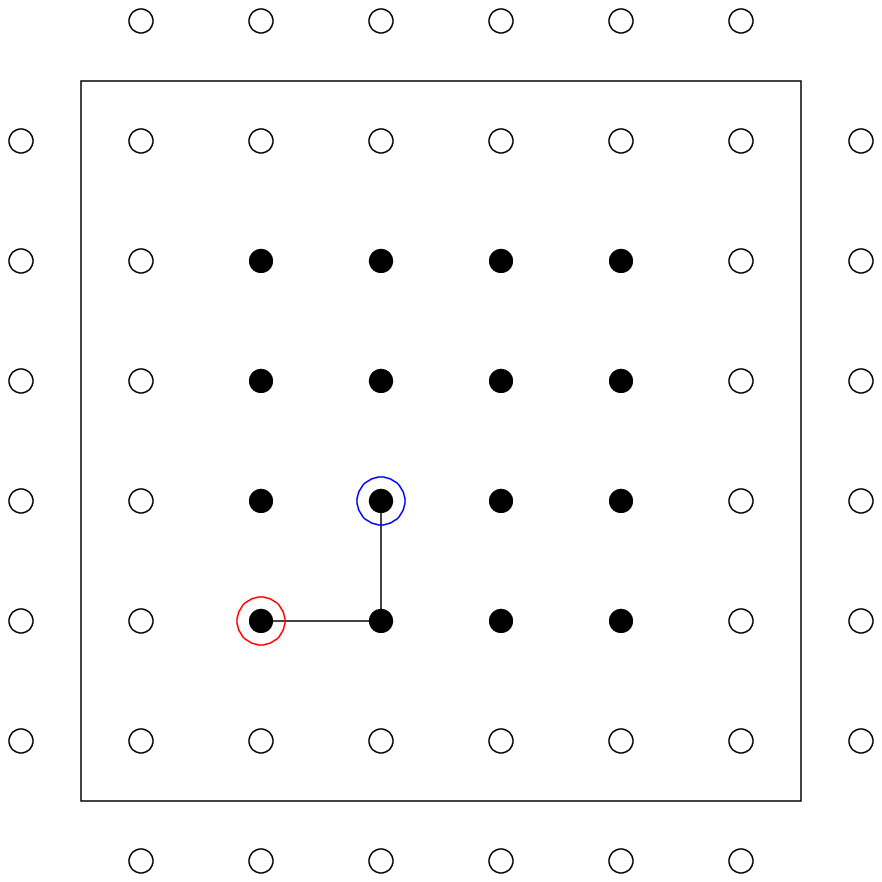}%
\end{center}
\end{minipage}
\hfill
\begin{minipage}[b]{0.47\linewidth}
\begin{center}
\includegraphics[width=3.5cm,angle=0,clip]{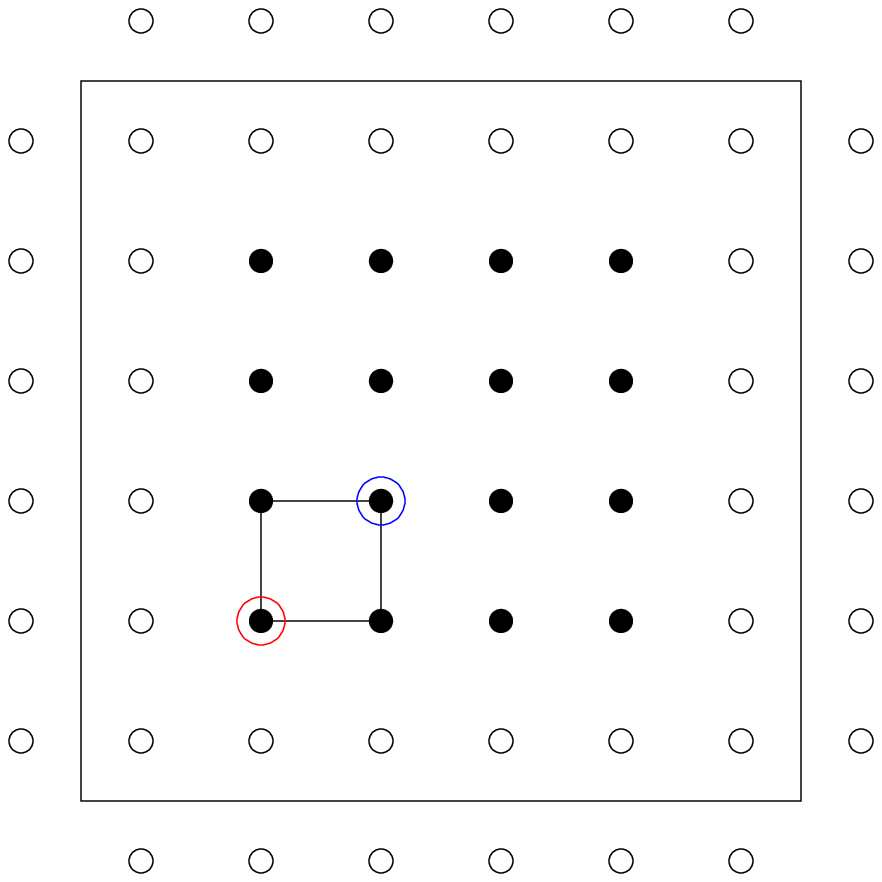}
\end{center}
\end{minipage}
\caption{The path of Loop3(left) and Loop4(right). The blue circle indicates the link upward and the red circle indicates the link downward.  }
\end{figure}
\begin{figure}[h]
\begin{minipage}[b]{0.47\linewidth}
\begin{center}
\includegraphics[width=3.5cm,angle=0,clip]{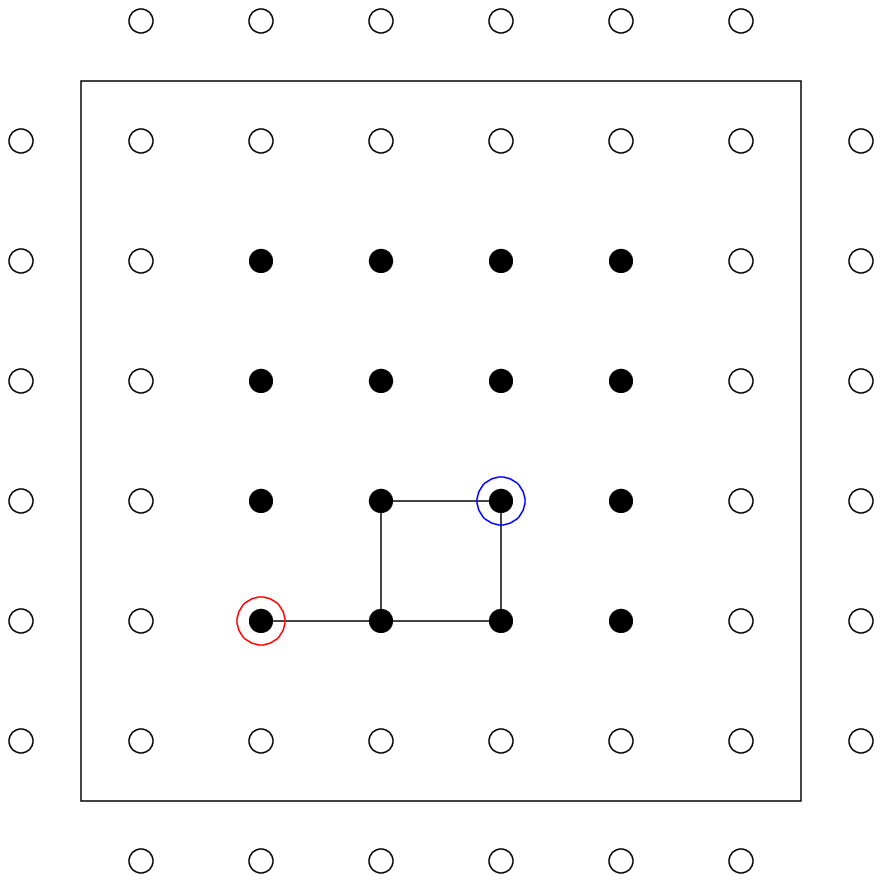}
\end{center}
\end{minipage}
\hfill
\begin{minipage}[b]{0.47\linewidth}
\begin{center}
\includegraphics[width=3.5cm,angle=0,clip]{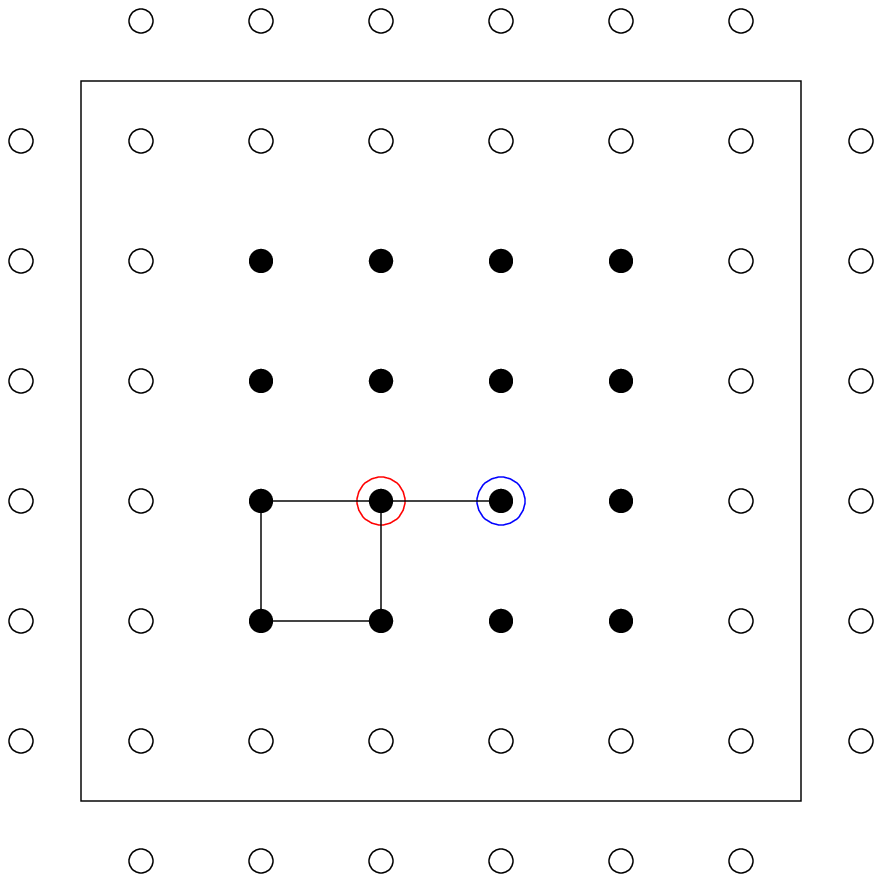}
\end{center}
\end{minipage}
\caption{The path of Loop7(left) and Loop8(right).   }
\end{figure}
\begin{figure}[h]
\begin{minipage}[b]{0.47\linewidth}
\begin{center}
\includegraphics[width=3.5cm,angle=0,clip]{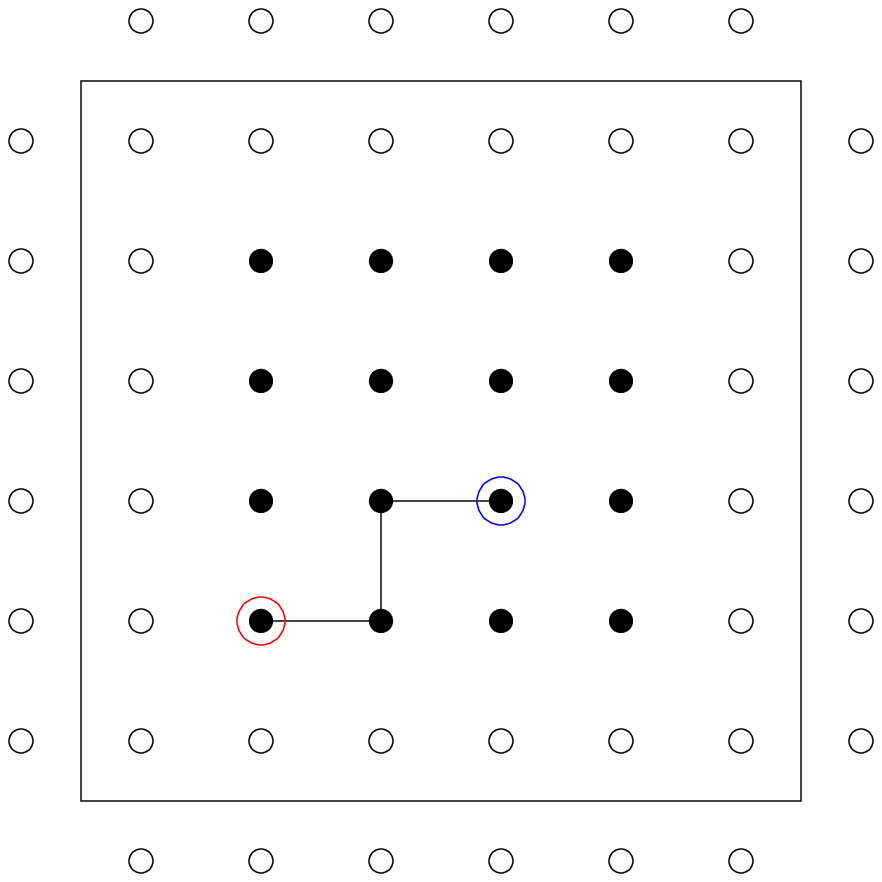}
\end{center}
\end{minipage}
\hfill
\begin{minipage}[b]{0.47\linewidth}
\begin{center}
\includegraphics[width=3.5cm,angle=0,clip]{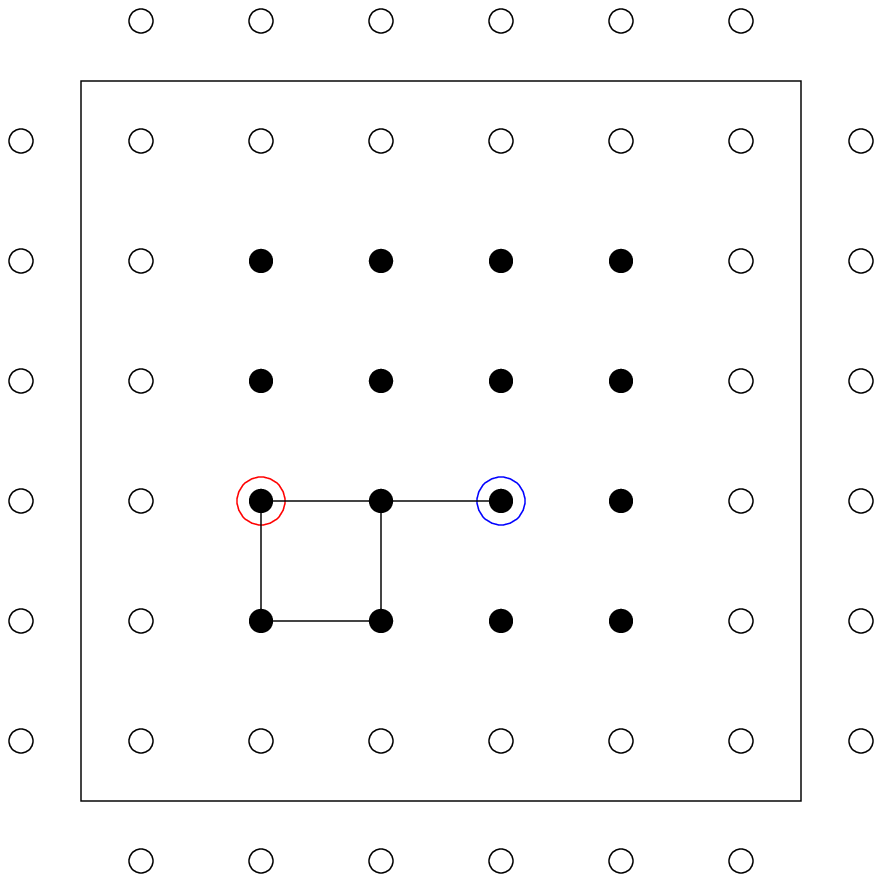}
\end{center}
\end{minipage}
\caption{The path of Loop9(left) and Loop10(right). .  }
\end{figure}
\begin{figure}[h]
\begin{minipage}[b]{0.47\linewidth}
\begin{center}
\includegraphics[width=3.5cm,angle=0,clip]{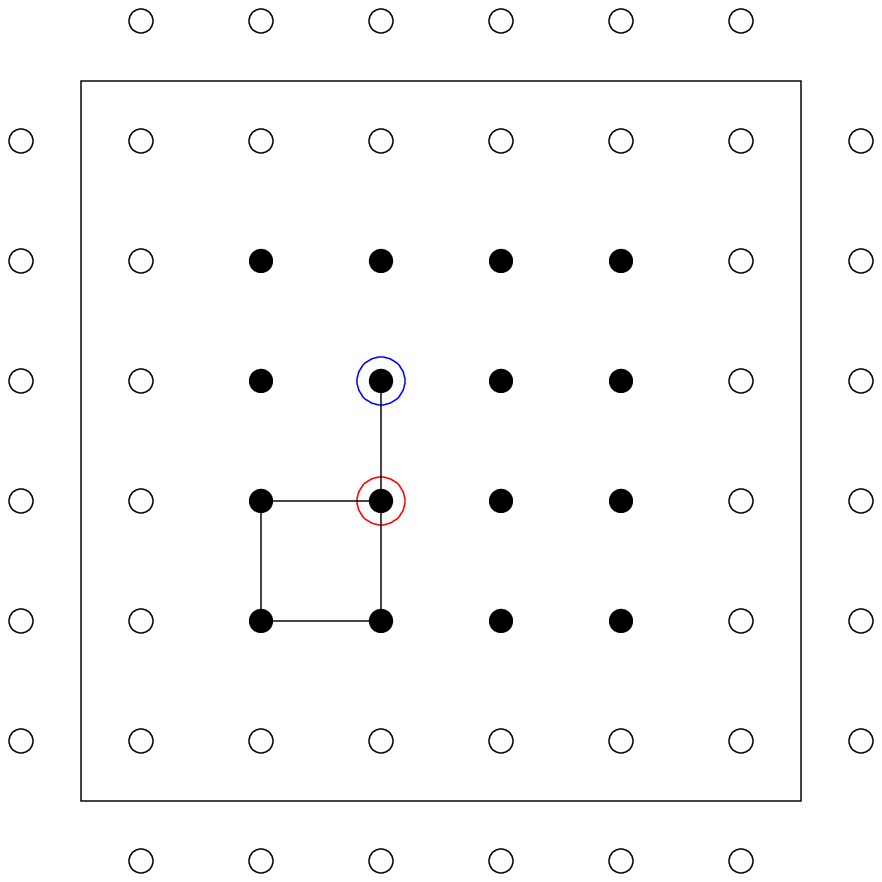}
\end{center}
\end{minipage}
\hfill
\begin{minipage}[b]{0.47\linewidth}
\begin{center}
\includegraphics[width=3.5cm,angle=0,clip]{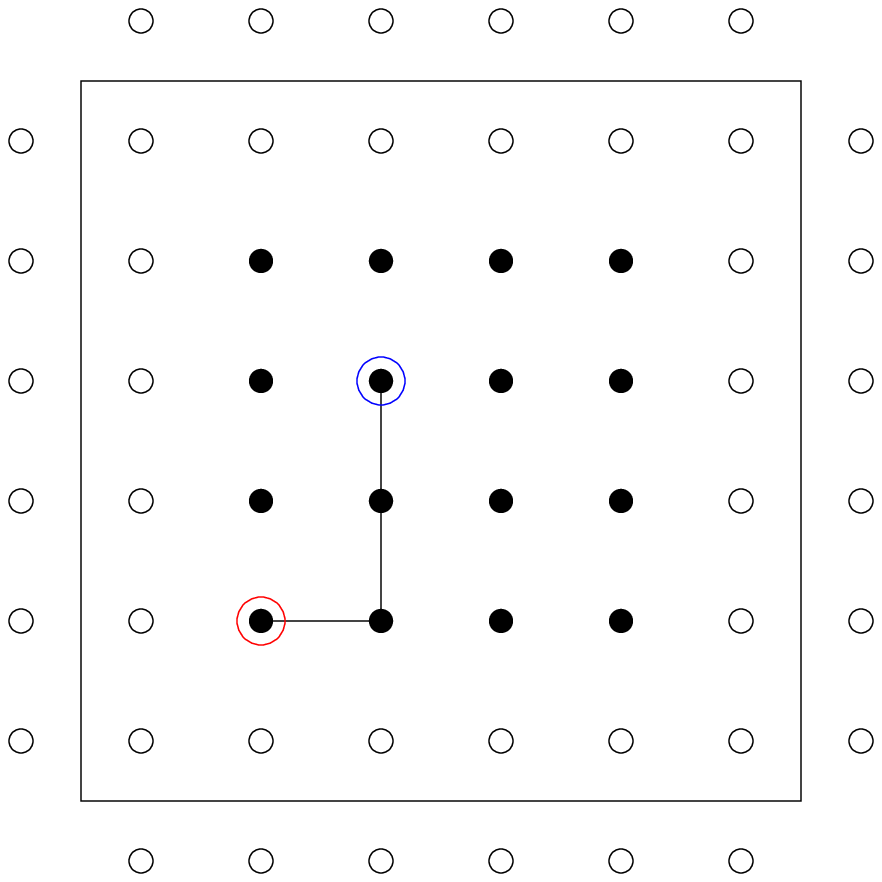}
\end{center}
\end{minipage}
\caption{The path of Loop13(left) and Loop14(right).   }
\end{figure}
\begin{figure}[h]
\begin{minipage}[b]{0.47\linewidth}
\begin{center}
\includegraphics[width=3.5cm,angle=0,clip]{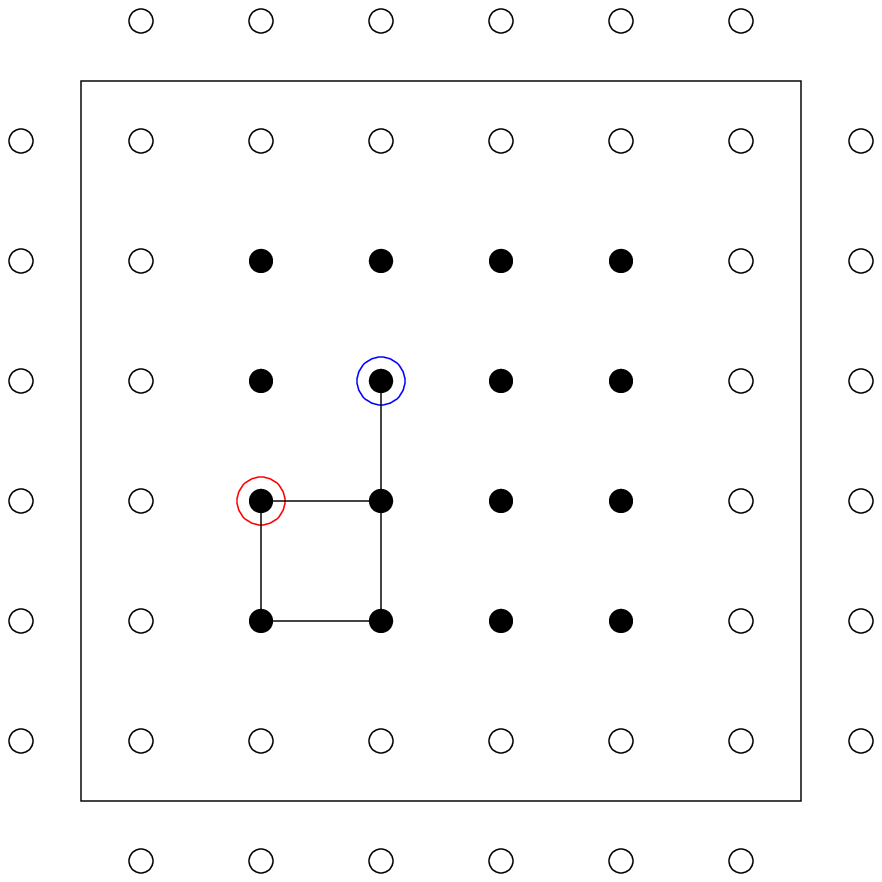}%
\end{center}
\end{minipage}
\hfill
\begin{minipage}[b]{0.47\linewidth}
\begin{center}
\includegraphics[width=3.5cm,angle=0,clip]{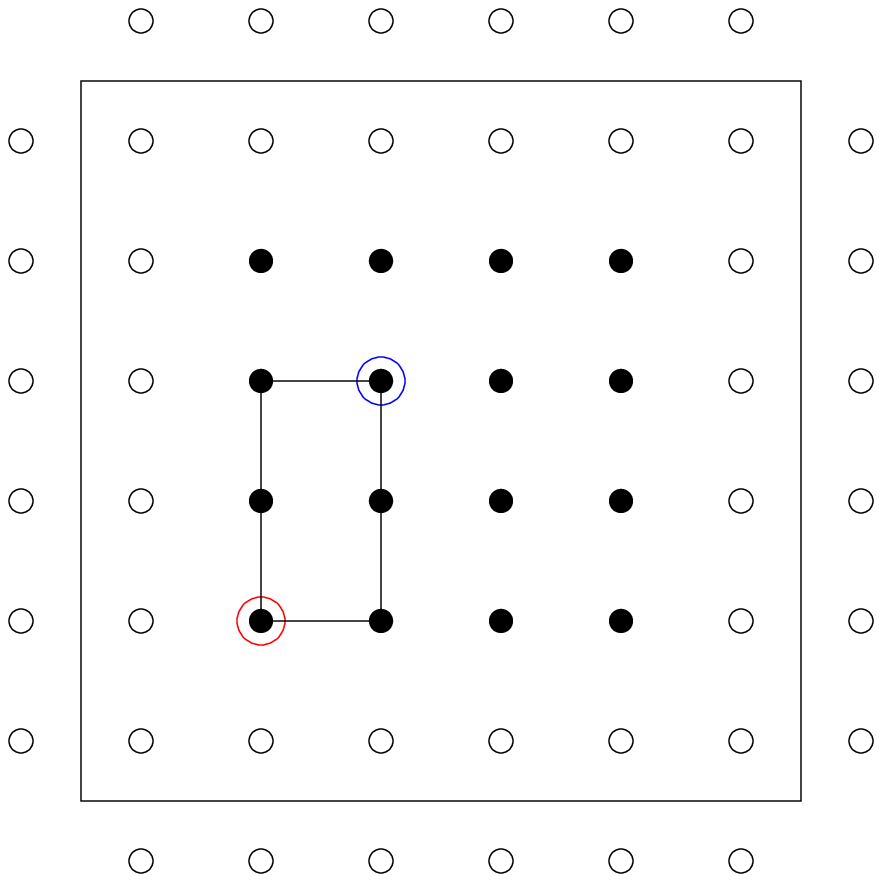}
\end{center}
\end{minipage}
\caption{The path of Loop15(left) and Loop16(right).   }
\end{figure}
\begin{figure}[ht]
\begin{minipage}[b]{0.47\linewidth}
\begin{center}
\includegraphics[width=3.5cm,angle=0,clip]{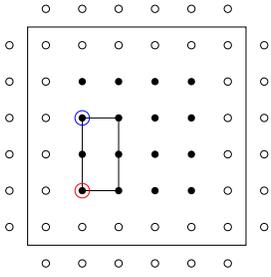} 
\end{center}
\end{minipage}
\hfill
\begin{minipage}[b]{0.47\linewidth}
\end{minipage}
\caption{The path of Loop17.  }
\end{figure}

When $z$ in DeGrand et al.\cite{DGHHN95} is replaced by $e_1\wedge e_2$ and the difference scale of $z$ and $t$ is ignored we obtain for the Loop 3
\begin{eqnarray}
&&L3[u_1,u_2]=t3[-\frac{1}{4}, u1, u2]\times t1[-\frac{1}{4}, u_1, u_2 + \frac{1}{4}]\nonumber\\
&& \times t2[-\frac{1}{4}, u_1, u_2 +\frac{1}{4}]\times t3[\frac{1}{4},  u_1 +\frac{1}{4}, u_2 + \frac{1}{4}]\nonumber\\
&& \times t2[\frac{1}{4}, u_1 + \frac{1}{4}, u_2]\times t1[\frac{1}{4}, u_1, u_2].\nonumber
\end{eqnarray}
\begin{figure}[htb]
\includegraphics[width=7cm,angle=0,clip]{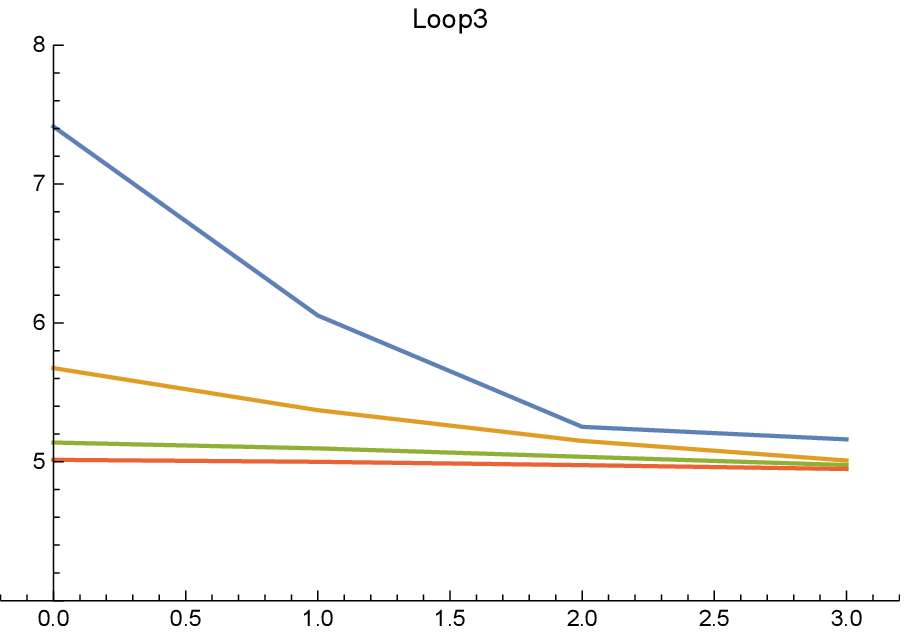}
\caption{ The absolute value of eigenvalues in Loop3 for a fixed $u_1$ as a function of $u_2$. ($\Delta u_i=1$ $(i=1,2)$)  }
\label{L3}
\end{figure}
\begin{figure}[h]
\begin{center}
\includegraphics[width=7cm,angle=0,clip]{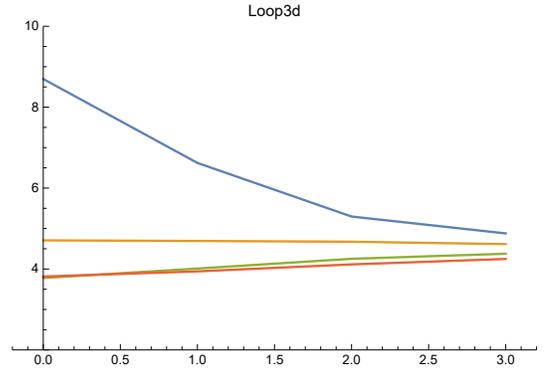} 
\end{center}
\caption{ The absolute value of eigenvalues in Loop3d (Double step length $\Delta t$.) }
\label{L3d}
\end{figure}

For the Loop 4
\begin{eqnarray}
&&L4[u_1,u_2]=t3[-\frac{1}{4}, u1, u2]\times t2[-\frac{1}{4}, u_1, u_2 + \frac{1}{4}]\nonumber\\
&&\times t1[-\frac{1}{4}, u_1+\frac{1}{4}, u_2 + \frac{1}{4}]\times t3\frac{1}{4}, u_1 + \frac{1}{4}, u_2 + \frac{1}{4}] \nonumber\\
&&\times t2[\frac{1}{4}, u_1 +\frac{1}{4}, u_2]\times t1[\frac{1}{4}, u_1, u_2].\nonumber
\end{eqnarray}
\begin{figure}[h]
\begin{center}
\includegraphics[width=7cm,angle=0,clip]{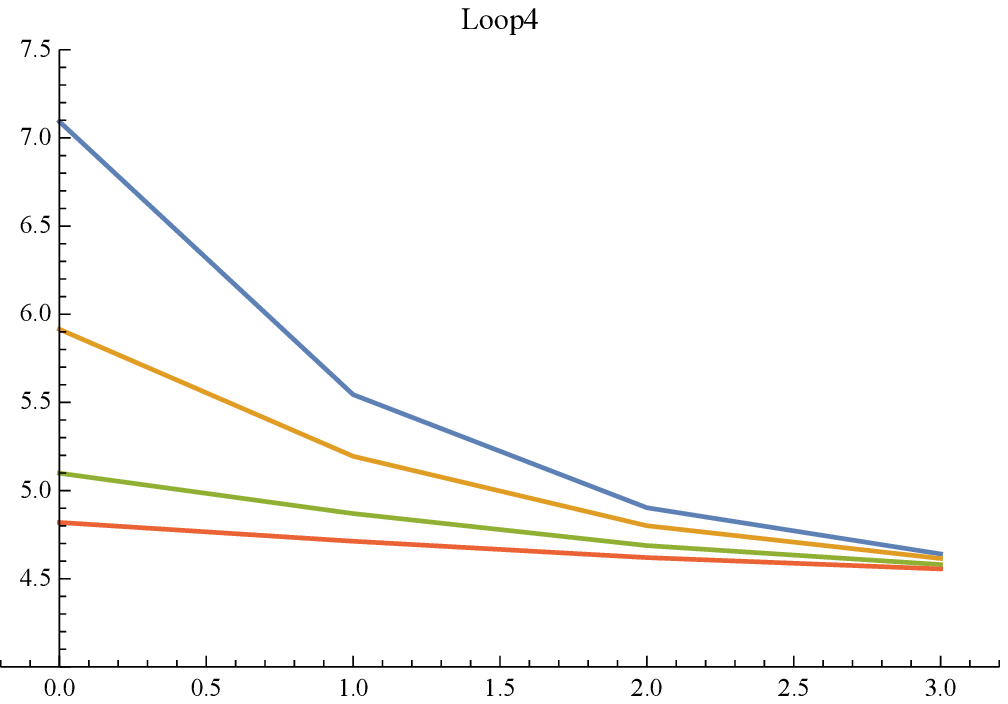}
\end{center}
\caption{ The absolute value of eigenvalues in Loop4 for a fixed $u_1$ as a function of $u_2$. ($\Delta u_i=1$ $(i=1,2)$).   }
\end{figure}

For the Loop7
\begin{eqnarray}
&&L7[u_1,u_2]= t3[-\frac{1}{4}, u_1, u_2]\times t1[-\frac{1}{2}, u_1+\frac{1}{2}, u_2] \nonumber\\
&&\times t2[-\frac{1}{4}, u_1 + \frac{1}{4}, u_2 + \frac{1}{4}]\nonumber\\
&&\times t3[\frac{1}{2}, u_1 + \frac{1}{4}, u_2 + \frac{1}{4}]\times t1[\frac{1}{4}, u_1 + \frac{1}{4},  u_2 + \frac{1}{4}]\nonumber\\
&&\times t2[\frac{1}{4}, u_1 +\frac{1}{4}, u_2]\times t1[\frac{1}{4}, u_1, u_2].\nonumber
\end{eqnarray}
\begin{figure}[h]
\begin{center}
\includegraphics[width=7cm,angle=0,clip]{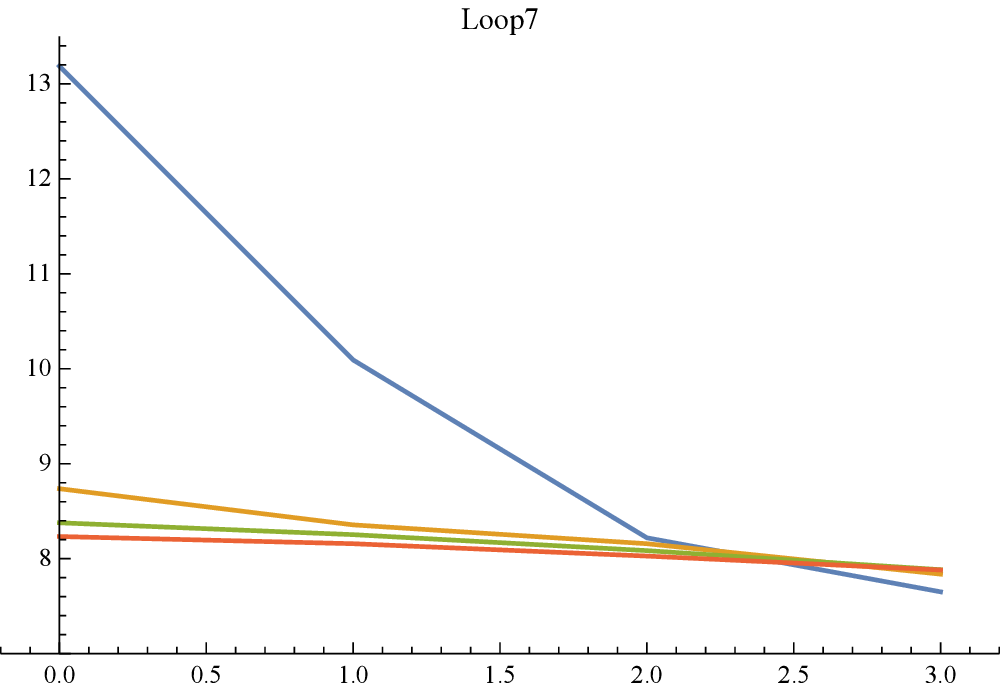}
\end{center}
\caption{ The absolute value of eigenvalues in Loop7 for a fixed $u_1$ as a function of $u_2$. ($\Delta u_i=1$ $(i=1,2)$)} 
\end{figure}

For the Loop8
\begin{eqnarray}  
&&L8[u_1,u_2]=t2[-\frac{1}{4}, u_1, u_2]\times t1[-\frac{1}{4}, u_1, u_2 + \frac{1}{4}]\nonumber\\
&&\times t3[-\frac{1}{4}, u_1+\frac{1}{2}, u_2]\times t2[-\frac{1}{4}, u_1 + \frac{1}{2}, u_2 + \frac{1}{4}] \nonumber\\
&&\times t3[\frac{1}{4}, u_1 + \frac{1}{2}, u_2 + \frac{1}{4}]\times t1[\frac{1}{4}, u_1 + \frac{1}{4}, u_2 + \frac{1}{4}]\nonumber\\
&&\times t2[\frac{1}{4}, u_1 + \frac{1}{4}, u_2]\times t1[\frac{1}{4}, u_1, u_2].\nonumber
\end{eqnarray}
\begin{figure}[h]
\begin{center}
\includegraphics[width=7cm,angle=0,clip]{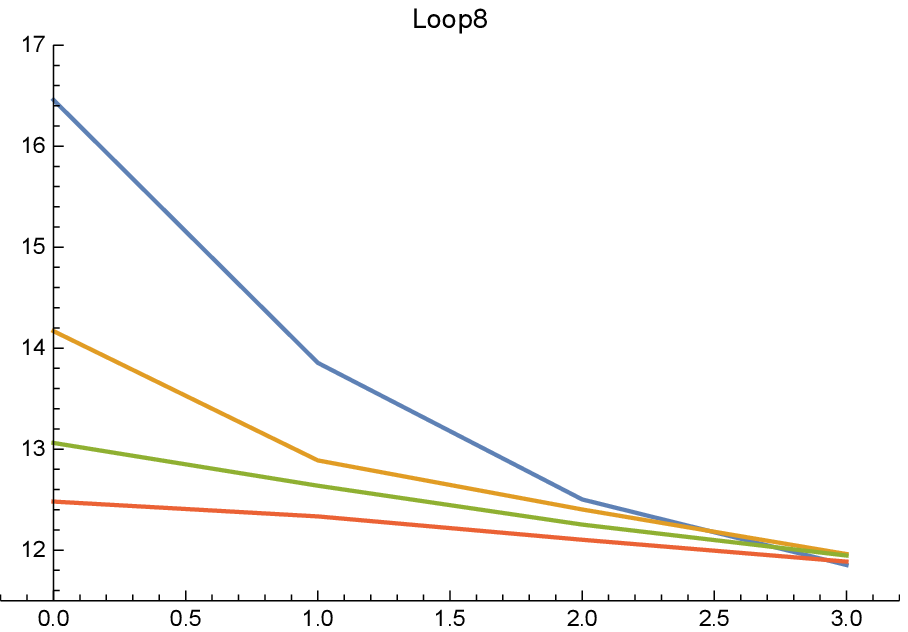}
\end{center}
\caption{ The absolute value of eigenvalues in Loop8 for a fixed $u_1$ as a function of $u_2$. ($\Delta u_i=1$ $(i=1,2)$)   }
\end{figure}

For the Loop9
\begin{eqnarray}
&&L9[u_1,u_2]=t3[-\frac{1}{4}, u_1, u_2]\times t1[-\frac{1}{4}, u_1, u_2]\nonumber\\
&&\times t2[-\frac{1}{4}, u_1, u_2 +\frac{1}{4}]\times t1[-\frac{1}{4},   u_1 + \frac{1}{4}, u_2 +\frac{1}{4}]\nonumber\\
&&\times t3[\frac{1}{4}, u_1 + \frac{1}{4}, u_2 + \frac{1}{4}]\times t1[\frac{1}{4}, u_1 + \frac{1}{4},   u_2 + \frac{1}{4}]\nonumber\\
&&\times t2[\frac{1}{4}, u_1 + \frac{1}{4}, u_2]\times t1[\frac{1}{4}, u_1, u_2].\nonumber
\end{eqnarray}  
\begin{figure}[h]
\begin{center}
\includegraphics[width=7cm,angle=0,clip]{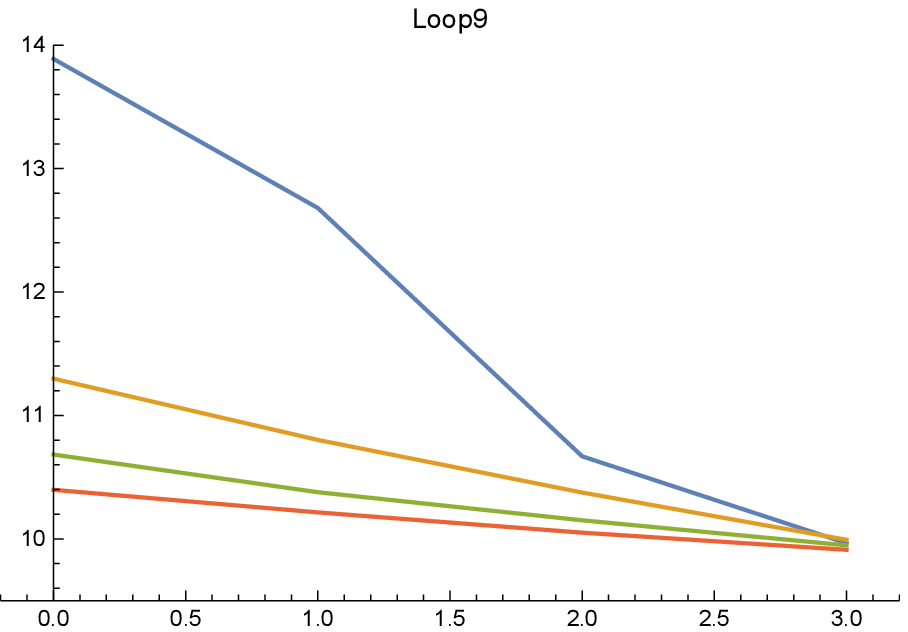}
\end{center}
\caption{ The absolute value of eigenvalues in  Loop9 for a fixed $u_1$ as a function of $u_2$. ($\Delta u_i=1$ $(i=1,2)$) }
\end{figure}

For the Loop10
\begin{eqnarray}
&&L10[u_1,u_2]=t2[-\frac{1}{4}, u_1, u_2 + \frac{1}{4}]\times t3[-\frac{1}{4}, u_1, u_2 + \frac{1}{4}]\nonumber\\
&&\times t1[-\frac{1}{2}, u_1 + \frac{1}{2},  u_2 + \frac{1}{4}]\times t3[\frac{1}{4}, u_1 + \frac{1}{2}, u_2 + \frac{1}{4}]\nonumber\\
&&\times t1[\frac{1}{4}, u_1 + \frac{1}{4}, u_2 +\frac{1}{4}]\times t2[\frac{1}{4}, u_1 + \frac{1}{4}, u_2]\times t1[\frac{1}{4}, u_1, u_2].\nonumber
\end{eqnarray}
\begin{figure}[h]
\begin{center}
\includegraphics[width=7cm,angle=0,clip]{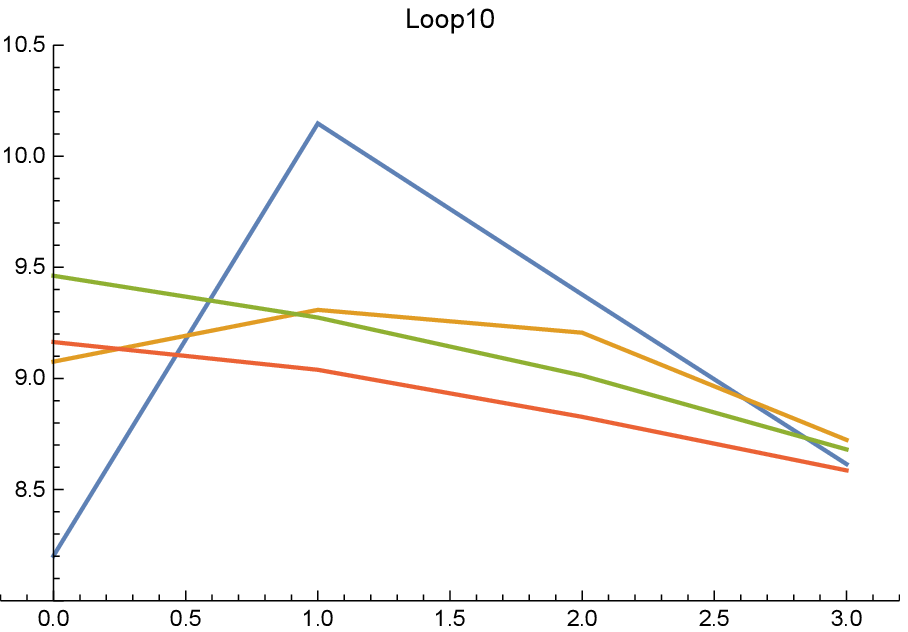}
\end{center}
\caption{ The absolute value of eigenvalues in Loop10 for a fixed $u_1$ as a function of $u_2$. ($\Delta u_i=1$ $(i=1,2)$) }
\end{figure}

 For the Loop13
\begin{eqnarray}  
&&L13[u_1,u_2]=t3[-\frac{1}{4}, u_1, u_2 +\frac{1}{4}]\times t1[-\frac{1}{4}, u_1 + \frac{1}{4}, u_2 + \frac{1}{4}]\nonumber\\
&&\times t3[-\frac{1}{4}, u_1 + \frac{1}{4},  u_2 + \frac{1}{4}]\times t2[-\frac{1}{4}, u_1 + \frac{1}{4}, u_2 + 1/2]\times\nonumber\\
&&\times t3[\frac{1}{4}, u_1 +\frac{1}{4}, u_2 +\frac{1}{2}]\times t2[\frac{1}{2}, u_1 +\frac{1}{4}, u_2]\times t1[\frac{1}{4}, u_1, u_2].\nonumber
\end{eqnarray}
\begin{figure}[htb]
\begin{center}
\includegraphics[width=7cm,angle=0,clip]{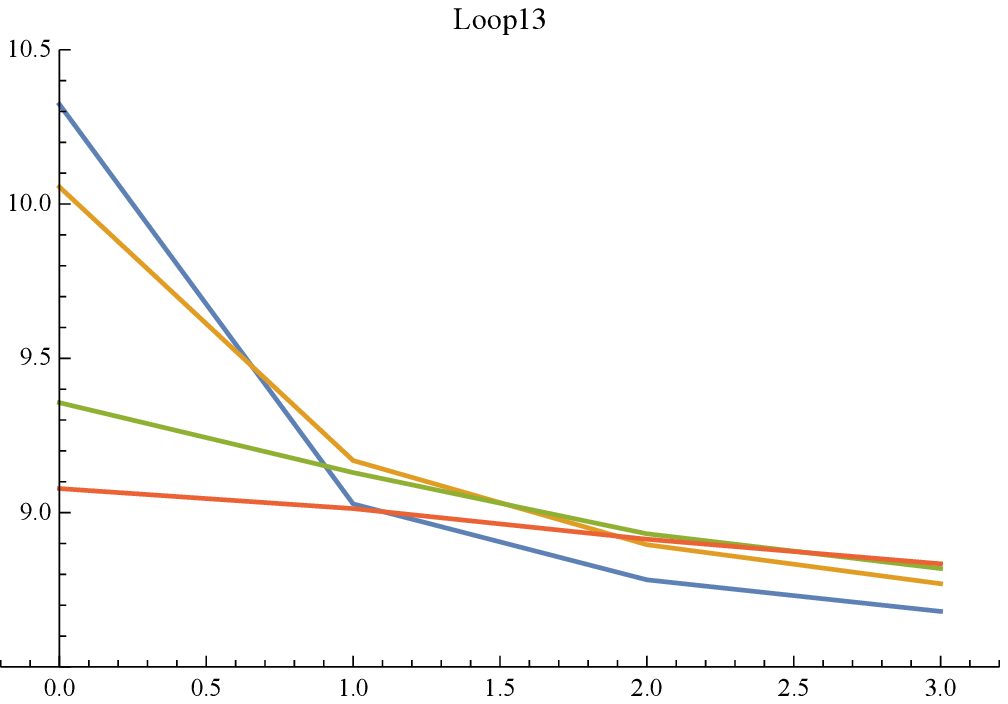}
\end{center}
\caption{ The absolute value of eigenvalues in Loop13 for a fixed $u_1$ as a function of $u_2$. ($\Delta u_i=1$ $(i=1,2)$) }
\end{figure}

 For the Loop14
\begin{eqnarray}
&&L14[u_1,u_2]=t3[-\frac{1}{4}, u_1, u_2 ]\times t1[-\frac{1}{4}, u_1 +\frac{1}{4}, u_2 ]\nonumber\\
&&\times t2[-\frac{1}{2}, u_1 + \frac{1}{4},  u_2 + \frac{1}{2}]\times t3[\frac{1}{4}, u_1 + \frac{1}{4}, u2 +\frac{1}{2}]\nonumber\\
&&\times t2[\frac{1}{2}, u_1 +\frac{1}{4}, u_2]\times t1[\frac{1}{4}, u_1, u_2].\nonumber
\end{eqnarray}
\begin{figure}[htb]
\begin{center}
\includegraphics[width=7cm,angle=0,clip]{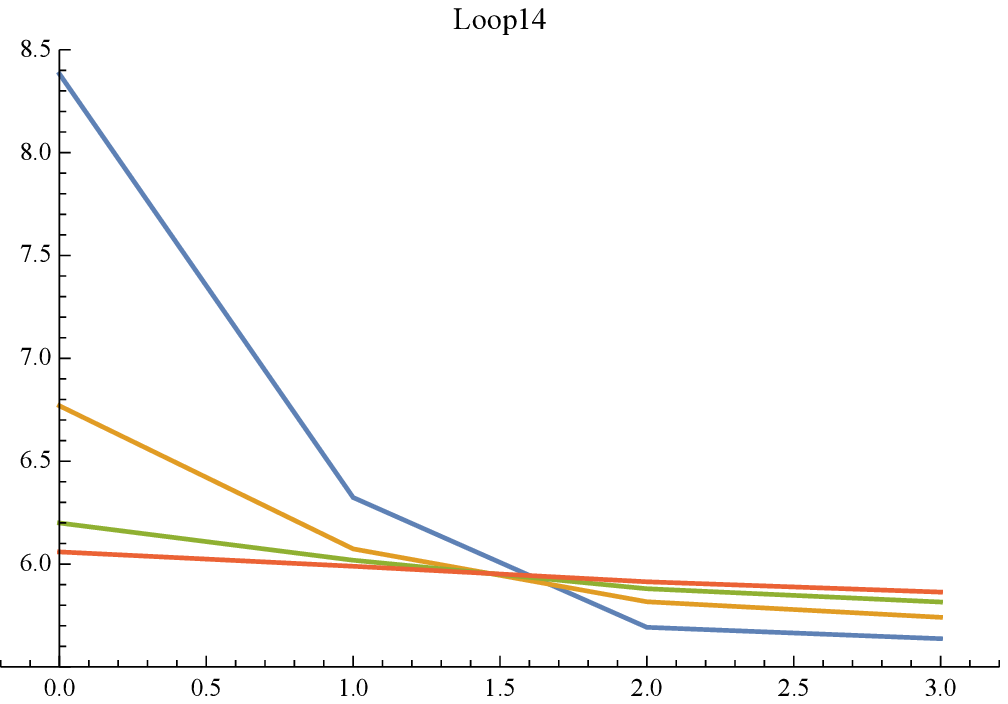}
\end{center}
\caption{ The absolute value of eigenvalues in Loop14 for a fixed $u_1$ as a function of $u_2$. ($\Delta u_i=1$ $(i=1,2)$) }
\end{figure}

 For the Loop15
\begin{eqnarray}
&&L15[u_1,u_2]=t2[-\frac{1}{4}, u_1, u_2 +\frac{1}{4}]\times t3[-\frac{1}{4}, u_1, u_2 + \frac{1}{4}]\nonumber\\
&&\times t1[-\frac{1}{4}, u_1 + \frac{1}{4},  u_2 + \frac{1}{4}]\times t2[-\frac{1}{4}, u_1 + \frac{1}{4}, u_2 +\frac{1}{2}]\nonumber\\
 &&\times t3[\frac{1}{4}, u_1 + \frac{1}{4},   u_2 +\frac{1}{2}]\times t2[\frac{1}{2}, u_1 + \frac{1}{4}, u_2]\times t1[\frac{1}{4}, u_1, u_2].\nonumber
\end{eqnarray}
\begin{figure}[htb]
\begin{center}
\includegraphics[width=7cm,angle=0,clip]{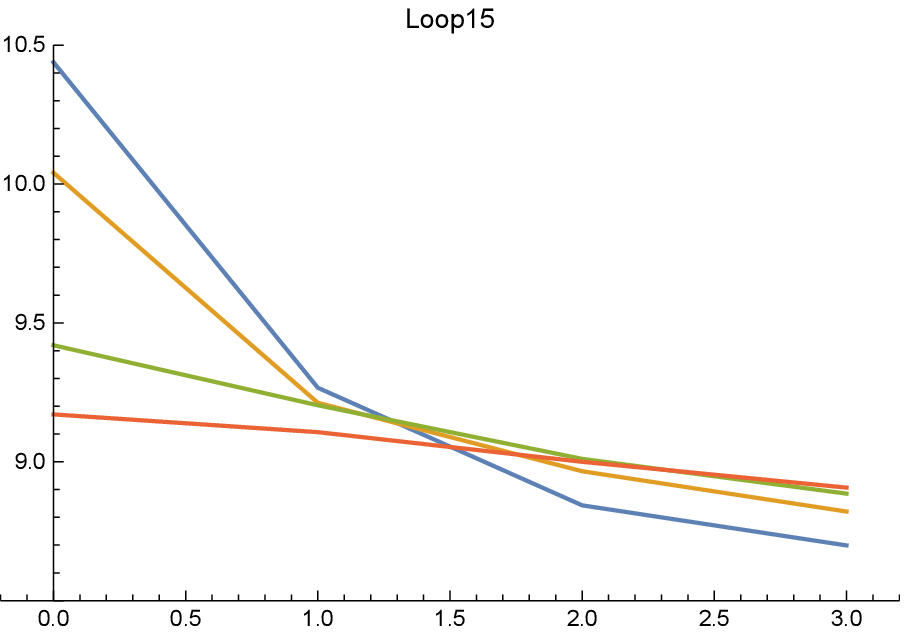}%
\end{center}
\caption{ The absolute value of eigenvalues in Loop15 for a fixed $u_1$ as a function of $u_2$. ($\Delta u_i=1$ $(i=1,2)$) }
\end{figure}

 For the Loop16
\begin{eqnarray}
&&L16[u_1,u_2]=t3[-\frac{1}{4}, u_1, u_2]\times t2[-\frac{1}{4}, u_1, u_2 + \frac{1}{4}]\nonumber\\
&&\times t1[-\frac{1}{2}, u_1 + \frac{1}{2},  u_2 +\frac{1}{4}]\times t3[\frac{1}{4}, u_1 + \frac{1}{2}, u_2 + \frac{1}{4}]\nonumber\\
&&\times t1[\frac{1}{4}, u_1 + \frac{1}{4},  u_2 + \frac{1}{4}]\times t2[\frac{1}{4}, u_1 + \frac{1}{4}, u_2]\times t1[\frac{1}{4}, u_1, u_2].\nonumber
\end{eqnarray}
\begin{figure}[ht]
\begin{center}
\includegraphics[width=7cm,angle=0,clip]{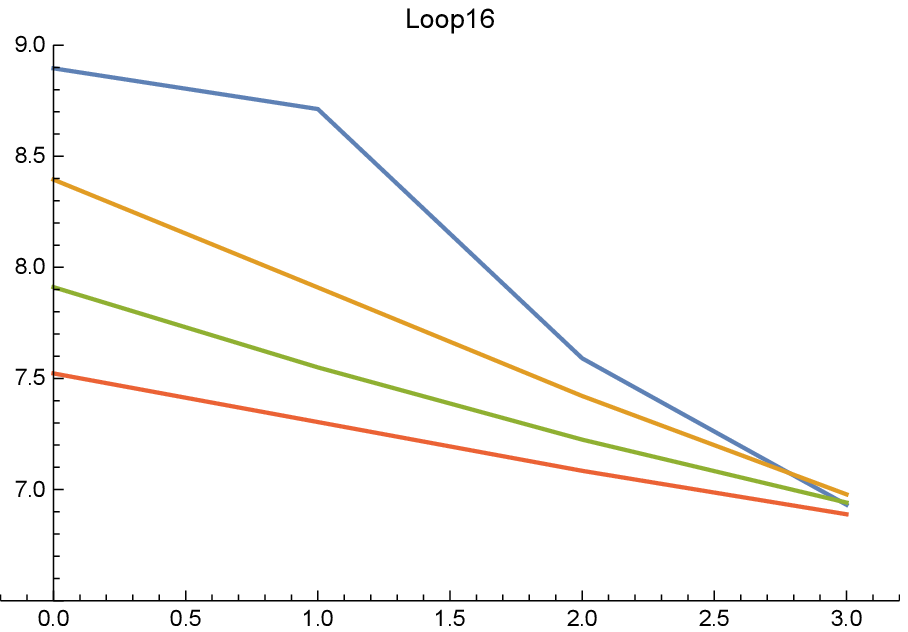}
\end{center}
\caption{ The absolute value of eigenvalues in Loop16 for a fixed $u_1$ as a function of $u_2$. ($\Delta u_i=1$ $(i=1,2)$) }
\end{figure}

 For the Loop17
\begin{eqnarray}
&&L17[u_1,u_2]=t3[-\frac{1}{4}, u_1 , u_2]\times t2[-\frac{1}{2}, u_1 , u_2 + \frac{1}{2}] \nonumber\\
&&\times t3[\frac{1}{4}, u_1 ,  u_2 + \frac{1}{2}]\times t1[-\frac{1}{4}, u_1 + \frac{1}{4}, u_2 +\frac{ 1}{2}]\nonumber\\
&&\times t2[\frac{1}{2}, u_1 + \frac{1}{4}, u_2]\times t1[ \frac{1}{4}, u_1, u_2].\nonumber
\end{eqnarray}
\begin{figure}[htb]
\begin{center}
\includegraphics[width=7cm,angle=0,clip]{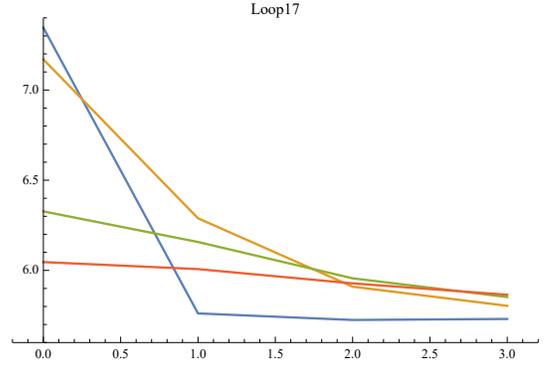} 
\end{center}
\caption{ The absolute value of eigenvalues in Loop17 for a fixed $u_1$ as a function of $u_2$. ($\Delta u_i=1$ $(i=1,2)$)}
\label{L17}
\end{figure}

The absolute value of the eigenvalues of $d{\bf S}[0, i]$ (Blue),$d{\bf S}[1,i]$ (Orange), $d{\bf S}[2,i]$ (Green), $d{\bf S}[3,i]$ (Red) for $i=0,1,2,3$ are plotted in Fig,\ref{L3} to Fig.30.

The scale of $e_3=e_1\wedge e_2$ cannot be fixed a priori. I took $\Delta e_1=\Delta e_2=1/4$, $\Delta e_3=1/2$.

I compared the eigenvalues of Loop3 and that of Loop3d, in which  $\Delta e_1=\Delta e_2=\Delta e_3=1/2$. 

The loop $L3[0,0]$ has the eigenvalue
\[
\chi(L3[0,0])=-5.33311 \pm 6.86919 \sqrt{-1}
\] 
and eigen-quaternion
\begin{eqnarray}
{\bf S}(L3[0,0])&=&0.594603\sqrt{-1} e_1+0.306216\sqrt{-1} e_0\nonumber\\
&&-0.743424\sqrt{-1}e_1\wedge e_2.\nonumber
\end{eqnarray}

The loop $L3d[0,0]$ has the eigenvalue
\[
\chi(L3d[0,0])=-6.37495 \pm 7.44031 \sqrt{-1}
\]
and the eigen-quaternion
\begin{eqnarray}
{\bf S}(L3d[0,0])&=&0.180682\sqrt{-1} e_1+0.263462\sqrt{-1} e_0\nonumber\\
&&-0.947598\sqrt{-1}e_1\wedge e_2.\nonumber
\end{eqnarray}

The eigenvalues $\chi(L3[i,j])$ $(0\leq i,j\leq 1)$ give the plaquette part of the Wilson loop.

Along the path $L3[0,0]$,
\begin{eqnarray}
 d{\bf S}(L3[0,0])&=&dS_1[0,0]+dS_2[\frac{1}{4},0]+dS_3[\frac{1}{4},\frac{1}{4}]\nonumber\\
  &&-dS_2[\frac{1}{4},\frac{1}{4}]-dS_1[\frac{1}{4},0]-dS_3[\frac{1}{4},0]
\end{eqnarray}
has a non-zero quaternion element in the right-upper corner of the $4\times 4$ matrix, and other elements are zero. 

 The sum over $d{\bf S}(L)$ corresponds to the loop part,
 and the sum over $\chi(L)$ corresponds to the interpolating surface part of the Wilson action\cite{Wilson74} in the lattice gauge theory.
 
 It can be checked by comparing the $\chi(L1[0,0]) \times 4\sim\chi(L28[0,0])=$, and  $d{\bf S}(L1[0,0])\times 2=d{\bf S}(L28[0,0]). $
 \newpage
\section{Recursive renormalization group analysis of $(2+1)D$ lattices}
We try to derive the effective action of the Weyl spinors by making the lattice spacing reduced by multilpying a factor $1/2$ at each step.

Migdal \cite{Migdal75a,Migdal75b} derived the scaling relation of the partition function in $D$ dimensional system
\begin{eqnarray}
&&Z_{2L}[{\bf S}[_\Gamma=\int\prod (d{\bf S}_{int}\prod_{i=1}^4 Z_{L}[{\bf S}_{\gamma i}]\nonumber\\
&&(d{\bf S})=d^n{\bf S} \delta(1-{\bf S}^2)\Gamma(n/2)\pi^{-n/2}\nonumber
\end{eqnarray}
where $[{\bf S}_{\gamma i} ]$ is the set of of spins on the boundaryof the $2L$-square, $[{\bf S}_{\gamma i}]$ is the sets of spins on the boundary of $L$-squares 
and $[{\bf S}_{int}]$ are the spins on the internal boundaries of the $L$-squares.

The Migdal-Kadanoff (MK) renormalization technique in ${\bf Z}^2$ model \cite{Creutz80,Migdal75a,Migdal75b,Kadanoff77} is to define the partition function $Z=Tr(T(\beta_0,\beta_1)^2)=Tr(T'(\beta_0',\beta_1'))$ that satisfy renormalization group equation. 

The parameters $\beta_0, \beta_1$ are obtained by solving
\begin{eqnarray}
&&T^2=[e^{\beta_0}\left(\begin{array}{cc}
e^{\beta_1}&e^{-\beta_1}\\
e^{-\beta_1}&e^{\beta_1}\end{array}\right)]^2=e^{\beta_0'}\left(\begin{array}{cc}
e^{\beta_1'}&e^{-\beta_1'}\\
e^{-\beta_1'}&e^{\beta_1'}\end{array}\right)\nonumber\\
&&=T'(\beta_0',\beta_1').\nonumber
\end{eqnarray}

A solution is 
\begin{eqnarray}
\beta_0&'=&\frac{1}{2}(4\beta_0+\log(4\cosh(2\beta_1)),\nonumber\\
\beta_1'&=&\frac{1}{2}\log(\cosh(2\beta_1)).\nonumber
\end{eqnarray}
 In $2D$ MK recursions, $\beta_1=({\log(1+{\sqrt 2}))}/{2}$ is a self dual point. 
 
The renormalization theory of quantized systems wasdiscussed by Gallavotti \cite{Gallavotti85}. A string Lagrangian and wave functions $e^{\sqrt{-1}(k x-{\mathcal E}(k)t)}$, where 
\begin{eqnarray}
&&{\mathcal E}(k)=\pm(\omega^2+c^2 k^2)^{1/2},\nonumber\\
&&p=\hbar k,\quad E=\hbar{\mathcal E}.\nonumber
\end{eqnarray}

The relation between the momentum $p$ and the velocity $v$ is
\begin{eqnarray}
&&v=\frac{d{\mathcal E}}{dk}=\frac{c^2 k}{(\omega^2+c^2 k^2)^{1/2}}\nonumber\\
&&p=v\frac{\omega\hbar /c^2}{(1-v^2/c^2)^{1/2}}.\nonumber
\end{eqnarray}

For lattice simulation of propagation of ultrasonic waves in $2D$ media, we adopt Luescher's domain decomposition method\cite{Luescher98a,Luescher98b,Luescher03}.

The choice of quaternion projective space on $2D$ planes is expected to reduce number of training parameters. Numerical calculation of the Generalized Conjugate Residual (GCR) method proposed by Luescher is to estimate the acceptance vs reject probability $P(C,C')$.

 The propagation in $e_1\wedge e_2$ should contain information on the effective mass of the Weyl spinor.
 
\section{Application of Luescher's Action to the Symplectic group $C\ell_{1,3}^+$}
\label{luescher}

In general, physical dynamics are represented by unitary group $U(n)$, which has bases of ${\bf C}$, orthogonal group $O(n)$, which has bases of ${\bf R}$, and the symplectic group $Sp(n)$, which has bases of ${\bf H}$\cite{Souriau70,Hestenes86,Porteous95,Lounesto01}.

In the Clifford algebra, transformation $X\to X'$ by a spin transformation 
$\left(\begin{array}{cc}
a& c\\
b& d\end{array}\right)$
is represented by\cite{Porteous95}
\[
\left(\begin{array}{cc}
a& c\\
b& d\end{array}\right)
\left(\begin{array}{cc}
x& x x^-\\
I& x^-\end{array}\right)
\left(\begin{array}{cc}
d^-& c^-\\
b^-&a^-\end{array}\right)=\lambda\left(\begin{array}{cc}
x'& x' x'^-\\
I& x'^-\end{array}\right),
\]
eigenvalues $\chi$ can be obtained from the $2\times 2$ matrix of the left-down corner of the $4\times 4$ matrix.

  We calculate $L[u_1,u_2]$ in $4\times 4$ matrices and pickup left-down corner $2\times 2$ complex matrices and calculates their eigen quaternions.
    
  Since the system is time-reversal symmetric, the two eigenvalues are conjugate with each other.   It is a characteristic for TR symmetric systems. 
  
The eigenvalues of the plaquett matrices corresponding to the Wilson loops on a planehave dependences on $[u_1,u_2]$ when it is close to $[0,0]$.
  \vskip 1 true cm
    
  Using the parametrization of ${\bf S}[u_1,u_2]$, we define
\[
\partial_1{\bf S}=\lim_{a\to 0}({\mathcal T}_{(1,0)}{\bf S}{{\mathcal T}_{(1,0)}}^{-1} -{\bf S})/a.
\]
Similarly, we define $\partial_2{\bf S}$.

\begin{figure}
\begin{minipage}[b]{0.47\linewidth}
\begin{center}
\includegraphics[width=4cm,angle=0,clip]{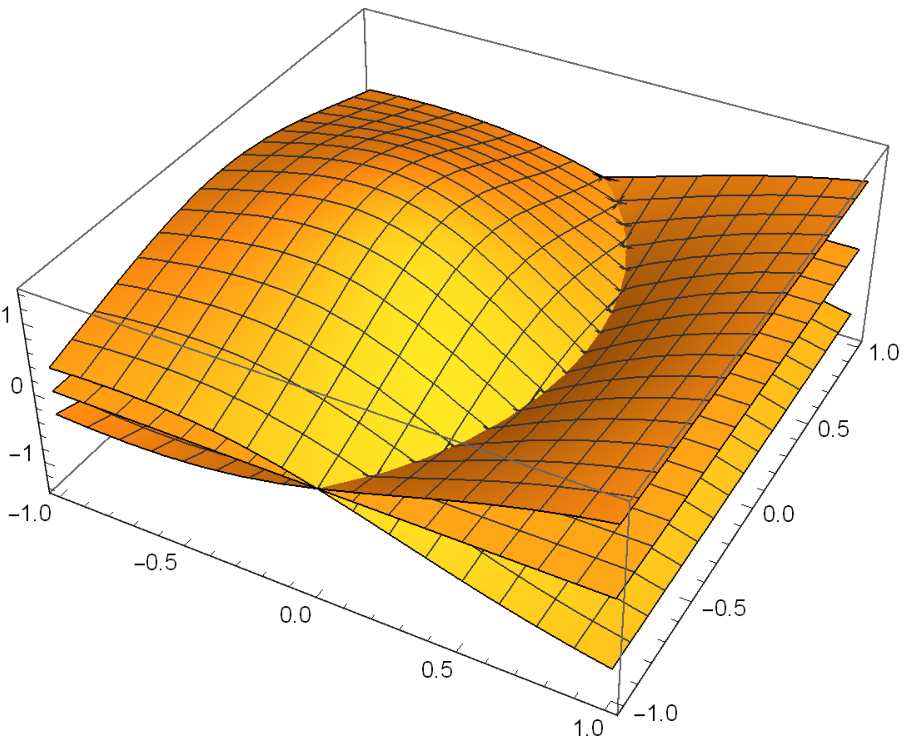}
\end{center}
\end{minipage}
\hfill
\begin{minipage}[b]{0.47\linewidth}
\begin{center}
\includegraphics[width=4cm,angle=0,clip]{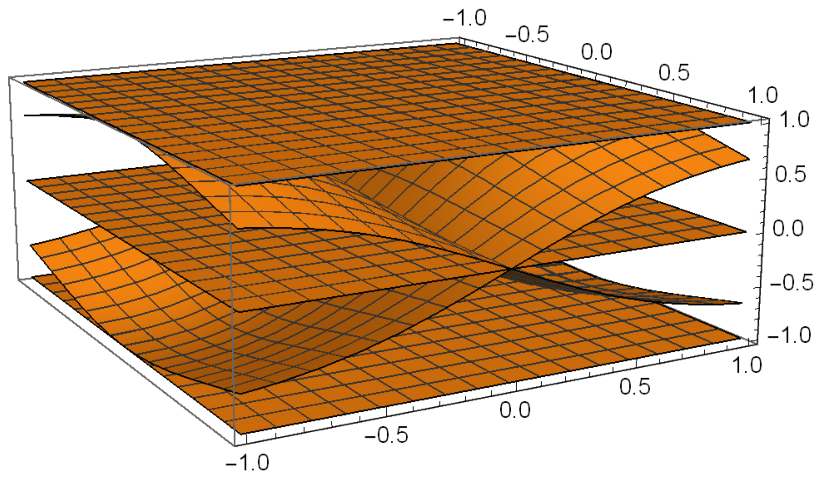}
\end{center}
\end{minipage}
\caption{The real part of $\partial_1{\bf S}[u_1,u_2]$ (left) and the imaginary part of $\partial_1{\bf S}[u_1,u_2]$ (right). }
\label{s1}
\begin{minipage}[b]{0.47\linewidth}
\begin{center}
\includegraphics[width=4cm,angle=0,clip]{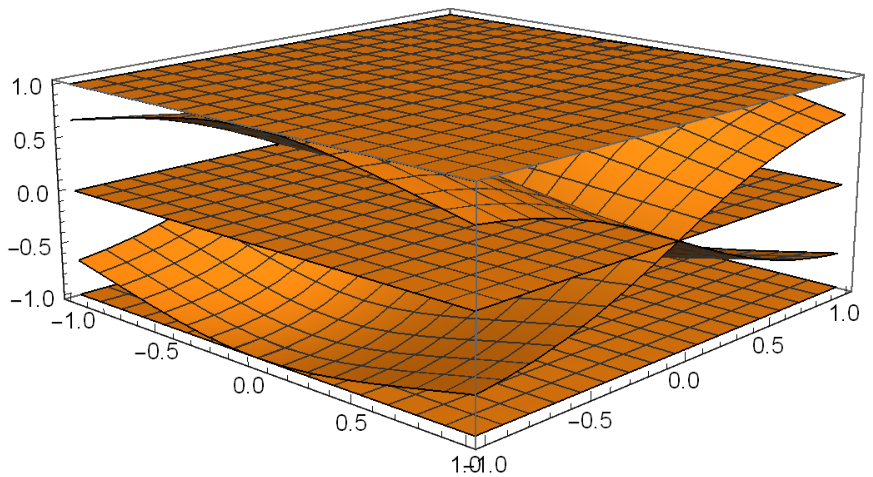}
\end{center}
\end{minipage}
\hfill
\begin{minipage}[b]{0.47\linewidth}
\begin{center}
\includegraphics[width=4cm,angle=0,clip]{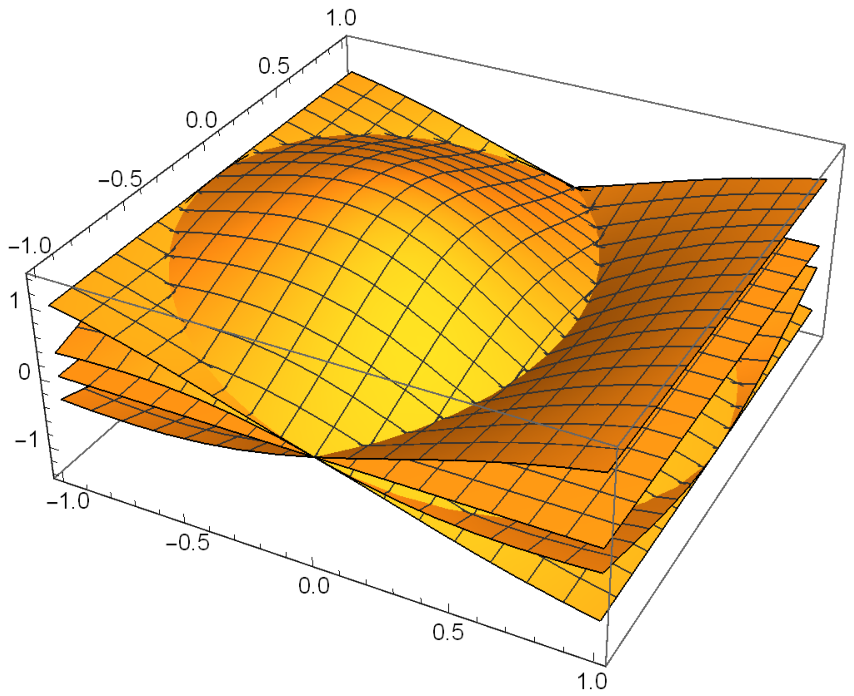}
\end{center}
\end{minipage}
\caption{The real part of $\partial_2{\bf S}[u_1,u_2]$ (left) and the imaginary part of $\partial_2{\bf S}[u_1,u_2])$. }
\label{s2}
\begin{minipage}[b]{0.47\linewidth}
\begin{center}
\includegraphics[width=4cm,angle=0,clip]{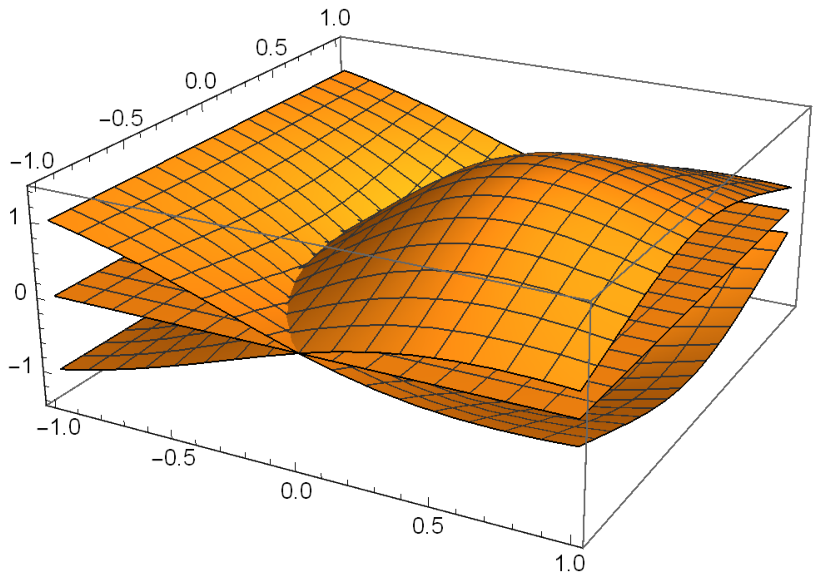} 
\end{center}
\end{minipage}
\hfill
\begin{minipage}{0.47\linewidth}
\begin{center}
\includegraphics[width=4cm,angle=0,clip]{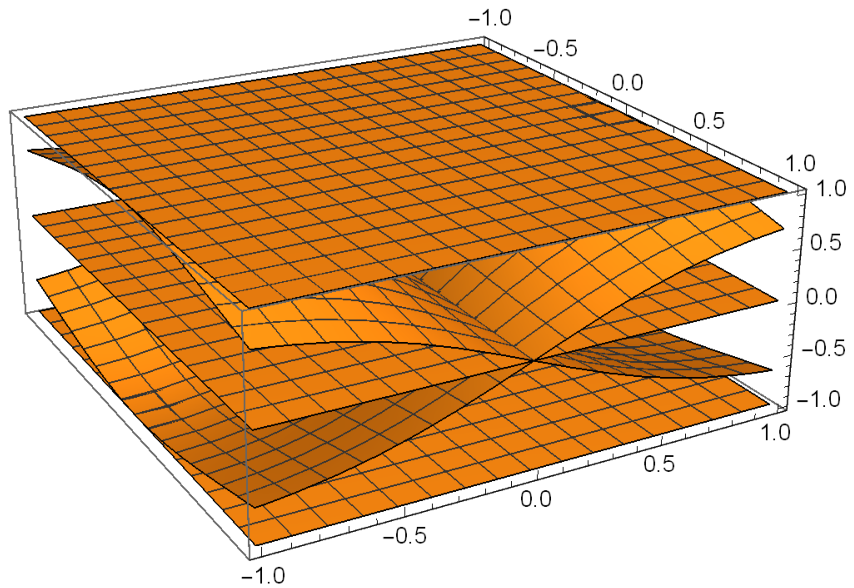}
\end{center}
\end{minipage}
\caption{The real part of $\partial_3{\bf S}[u_1,u_2]$ (left) and the imaginary part of $\partial_3{\bf S}[u_1,u_2]$ (right). }
\label{s3}
\begin{center}
\includegraphics[width=4cm,angle=0,clip]{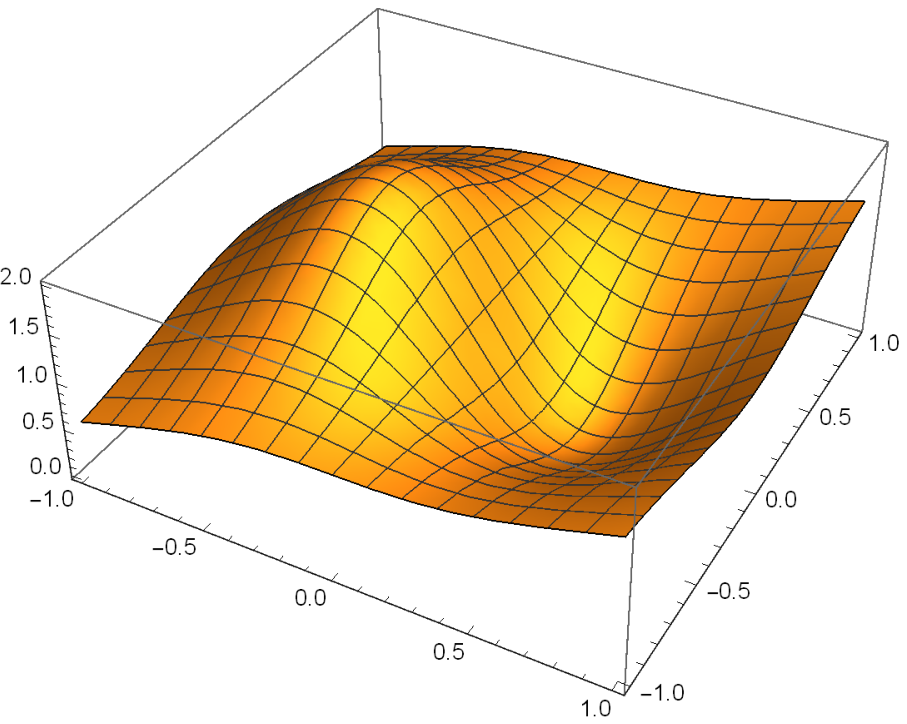}
\end{center}
\caption{ $\partial_i {\bf S}\cdot\partial_i {\tilde {\bf S}} [u_1,u_2]$  $(i=1,2)$}
\label{s12}
\begin{center}
\includegraphics[width=4cm,angle=0,clip]{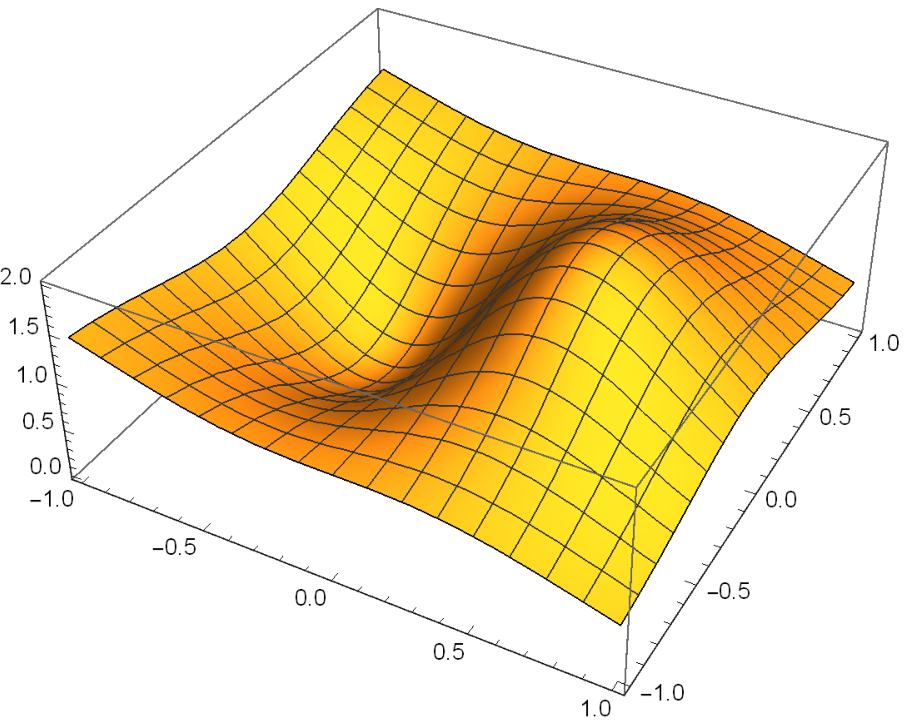}
\end{center}
\caption{$\partial_3{\bf S}\cdot\partial_3 {\tilde{\bf S}}[u_1,u_2]$.}
\label{ss3}
\end{figure}

The real part and the imaginary part of $\partial_i{\bf S}(u_1,u_2)$ $(i=1,2)$ in the region $-1\leq u_1\leq 1$ and $-1\leq u_2\leq 1$ are plotted in Fig. \ref{s1} and Fig. \ref{s2}. 

The wedge product $e_1\wedge e_2=e_3$ yields 
\[
\partial_3{\bf S}[u_1,u_2]=\lim_{a\to 0}({\mathcal T}_{(1,1)}{\bf S}{{\mathcal T}_{(1,1)}}^{-1} -{\bf S})/a.
\]
Its real and imaginary part are plotted in Fig.\ref{s3}.

 The action in Clifford algebra is a discrete sum \\
$\sum\partial_\mu{\bf S}(u_1,u_2)\cdot\partial_\mu{\tilde{\bf S}}(u_1,u_2)$, for $-1\leq u_1,u_2\leq1$ with an appropriate measure. 

$\partial_i {\bf S}\cdot\partial_i{\tilde{\bf S}}[u_1,u_2]$ for $i=1,2$ is 
{\small
\begin{eqnarray}
&&\partial_i {\bf S}\cdot \partial_i{\tilde {\bf S}}[u_1 u_2]=\nonumber\\
&&\frac{4{u_1}^3+{u_1}^4+4u_1(-1+{u_2}^2)+2{u_1}^2(1+{u_2}^2)+(1+{u_2}^2)^2}{(1+|u|^2)^2},\nonumber
\end{eqnarray}
}
which is shown in Fig.\ref{s12}.

 For $i=3$ 
{\small
\begin{eqnarray}
&&\partial_3 {\bf S}\cdot \partial_3{\tilde {\bf S}}[u_1 u_2]=\nonumber\\
&&\frac{-4{u_1}^3+{u_1}^4-4u_1(-1+{u_2}^2)+2{u_1}^2(1+{u_2}^2)+(1+{u_2}^2)^2}{(1+|u|^2)^2}.\nonumber
\end{eqnarray}
}
is shown in Fig.\ref{ss3}.

\begin{figure}[ht]
\begin{center}
\includegraphics[width=7cm,angle=0,clip]{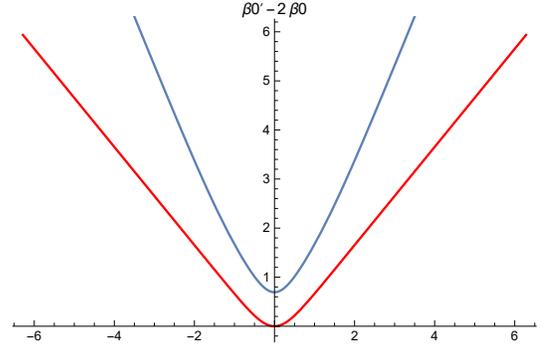}%
\end{center}
\caption{ $\beta_0'-2\beta_0$ as a function of $\beta_1$ of Clifford algebra. At $\beta_1=0$, it has the minimum $\log 2$. $\beta_0'-2\beta_0$ in MK method is indicated by red line.}
\label{beta0}
\end{figure}
\begin{figure}[htb]
\begin{center}
\includegraphics[width=7cm,angle=0,clip]{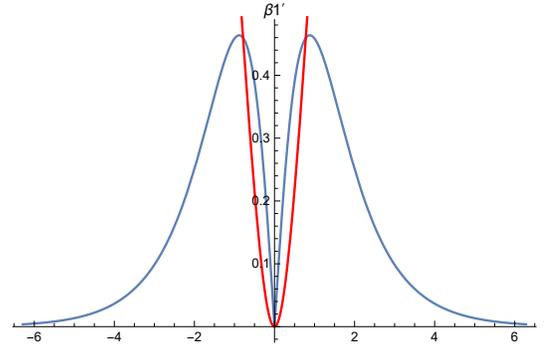}
\end{center}
\caption{ $\beta_1'/\sqrt{-1}$ as a function of $\beta_1$ in Clifford algebra. \\ At  $\beta_1=\pm\log(1+{\sqrt 2})=\pm 0.881374$ it has the maximum 0.463648.  $\beta_1'$ in MK method is indicated by red line}
\label{beta1}
\end{figure}

 In the Clifford algebra, the MK renormalization is modified to the equation 
 \[
 2T^2=T'(\beta_0',\beta_1').
 \]
 The difference of a factor 2 in the left hand side of the equation from that of MK is due to the choice of normalization $\sum_s 1=2$ \cite{Creutz80}.

Since for ${\bf a},{\bf b}\in {\bf H}$, ${\bf ab}=-{\bf a}\cdot{\bf b}+{\bf a}\wedge{\bf b}$, there appears a factor of $\sqrt{-1}$ difference in $\beta_1'$.
\begin{eqnarray}
T&=& e^{\beta_0}( \cosh(\beta_1)e_0+\sinh(\beta_1)e_1\wedge e_2)\nonumber\\
&=&e^{\beta_0}\left(\begin{array}{cc}
\cosh(\beta_1)&\sqrt{-1}\sinh(\beta_1)\\
\sqrt{-1}\sinh(\beta_1)&\cosh(\beta_1)\end{array}\right).\nonumber
\end{eqnarray}

\begin{eqnarray}
\beta_0'&=&\log [2 e^{2\beta_0}(\frac{\cosh(\beta_1)^4}{\cosh(\beta_1)^4+\sinh(\beta_1)^2}\nonumber\\
&&+\frac{\sinh(\beta_1)^2}{\cosh(\beta_1)^4+\sinh(\beta_1)^2})],\nonumber\\
\beta_1'&=&\cosh^{-1}(\frac{\cosh(\beta_1)^2}{\cosh(\beta_1)^4+\sinh(\beta_1)^2}).\nonumber
\end{eqnarray}

$\beta_0$ and $\beta_1/\sqrt{-1}$ in the Clifford algebra and corresponding $\beta_0$ and $\beta_1$ in the Unitary algebra are shown in Fig,\ref{beta0} and Fig.\ref{beta1}.

\section{Renormalization group and the scaling problem}
In order to study propagation of a phonon or a poson in fermions filled up to Fermi momentum $p_F$, renormalization group approach of Benfatto and Gallavotti\cite{BG95} is useful.

Gallavotti\cite{Gallavotti85} studied a specific Hamiltonian
\begin{eqnarray}
H_{quantum}&=&-\frac{\hbar^2}{{2\mu a^D}^2 }\sum_{ {\bar n a}\in\lambda_0}\frac{\partial^2}{\partial\varphi_{\bar n a}}
\nonumber\\
&&+\mu\frac{a^D}{2}\sum_{\bar n a\in\Lambda_0}[c^2\sum_{j=1}^D(\varphi_{\bar n a+\bar\epsilon_j a}-\varphi_{\bar n a})^2/a^2\nonumber\\
&&+(m_0c^2/\hbar)^2\varphi_{\bar n s}^2+I(\varphi_{\bar n a})
\end{eqnarray}
where $I(\varphi_{\bar n a})$ is a function of $\varphi_{\bar n a}$ bounded below.

On the $L_2$ space with the Gaussian measure
\[
''d\varphi''=\prod_{\bar x\in\lambda_0}d\varphi(\bar x),
\]
is
\[
T_t=exp[-(H_{quantum}-E)t/\hbar],\quad t\geq 0
\]
one defines $e(\bar\varphi)=$ ground state wave function for $H_{\rm quantum}$ and put a suffix $0$ for values for $t=0$, eg $H_0=H_{\rm quantum}|_{t=0}$.

$T_t^0(\bar\varphi,\bar\varphi')={\rm kernel\, of\,} T_t {\rm on}\quad L_2[\prod_{\bar n a} d\varphi_{\bar n a}]$.

Multiscale decomposition of the system is performed as
\[
C_{\xi,\eta}=\sum_{\bar n\in Z^D}\bar C_{\bar x+\bar n L,t}(\bar y,t'),
\]
where $\xi=(\bar n a,t)\in\Lambda_0\times S^1$ and $\eta=(\bar m a,t')$. We replace ${\bf R}$ of\cite{Gallavotti85} to $S^1$\cite{SF20} for the lattice simulation.

The measure $P(d\varphi)$ becomes
\begin{eqnarray}
P(d\varphi)&=& Z^{-1}[exp\{-\frac{\mu b a^D}{2\hbar}\sum_{\bar n a\in\lambda_0}\sum_m[\{(\varphi_{\bar n a, m b}-
\varphi_{\bar n  a,m b+b})^2/b^2\nonumber\\
&&+c^2\sum_{j=1}^D(\varphi_{\bar n a,m b}-\varphi_{\bar n a+\bar e_j a,mb})^2/a^2\}\nonumber\\
&&+(m_0c^2/\hbar)^2\varphi_{\bar n a, mb}+I(\varphi_{\bar n a, mb})\}]\prod_{\bar n, m}d\varphi_{\bar n, mb}\nonumber
\end{eqnarray}

One defines the measure
\begin{eqnarray}
P^{(\leq N)}(d\varphi)&=&Z_{N,a}^{-1}[exp\{-\frac{\mu a^d}{2c\hbar}\sum_\xi I(\varphi_\xi^{(\leq N)} \}]\nonumber\\
&&\times\prod_{j=0}^N P(d\varphi^{(j)}),
\end{eqnarray}

The interaction $I(\varphi)$ is assumed to have the form
\begin{eqnarray}
V(\varphi^{(\leq N)},\bar \lambda,N)&=&\sum_{\alpha=1}^t \lambda^{(\alpha)}\int_\Lambda v_N^{\alpha}(
\varphi_\xi^{(\leq N)},\partial\varphi_\xi^{(\leq N)})d\xi
\end{eqnarray}
in the continuum.

For all $N'\leq N$,
\begin{eqnarray}
&&\int_\Lambda v_{N'}^{(\alpha)} (\varphi_\xi^{(\leq N')}, \partial\varphi_\xi^{(\leq N')}d\xi\nonumber\\
&&=\int P(d\varphi^{(N'+1)})\cdots P(d\varphi^{(N)})\int_\Lambda v_N^{(\alpha)}(\varphi_x^{(\leq N)},\partial\varphi_\xi^{(\leq N)})d\xi.\nonumber\\
\end{eqnarray}

The effective interaction of the fields $\varphi^{(0)},\cdots,\varphi^{(k)}$ on their own length scale $\gamma^{-k} m^{-1}$ is
\begin{eqnarray}
e^{V^{(k)}, \varphi^{(\leq k)}}&=&\int exp[V(\varphi^{(\leq N)}] \nonumber\\
&&\times P(d\varphi^{(N)})\cdots P(d\varphi^{(k+1)}).
\end{eqnarray}

One defines the expectation value with respect to a probability measure as ${\mathcal E}( )$, the expectation value with respect to the Gaussian measure $P(d\varphi^{(k)})$ as ${\mathcal E}_k( )$, and $p$ random variables $x_1,\cdots,x_p$ of orders $n_1,\cdots,n_p$.

The truncated expectations of $x_1,\cdots,x_p$ of orders $n_1,\cdots,n_p$ are
\begin{eqnarray}
&&{\mathcal E}^T(x_1,\cdots,x_p; n_1,\cdots,n_p)\nonumber\\
&&=\frac{\partial^{n_1+\cdots+n_p}}{\partial \lambda_1^{n_1}\cdots\partial \lambda_p^{n_p}}\log {\mathcal E}(
e^{\lambda_1 x_1+\cdots+\lambda_p x_p})|_{\lambda_1=\cdots=\lambda_p=0}\nonumber\\
\end{eqnarray}

For $\omega_,\cdots, \omega_p \in {\bf R}$
\begin{eqnarray}
&&{\mathcal E}^T(\omega_1 x_1+\cdots+\omega_p x_p; n)\nonumber\\
&&=\sum_{n_1,\cdots,n_p | n_1+\cdots +n_p=n}\frac{n!\omega_1^{n_1}\cdots\omega_p^{n_p}}{n_1!\cdots n_p!}\nonumber\\
&&\quad\quad\times {\mathcal E}^T(x_1,\cdots,x_p; n_1,\cdots,n_p).\nonumber
\end{eqnarray}

When one ignores the convergence conditions
\begin{eqnarray}
\int P(d\varphi^{(N)}) e^V&\equiv& exp[\sum_{n=1}^\infty \frac{{\mathcal E}_N^T(V; n)}{n\!}]\nonumber\\
&=&exp[V^{(N-1)}],\nonumber
\end{eqnarray}
where ${\mathcal E}^T(V;n)={\mathcal E}^T(x_1,\cdots,x_p; n_1,\cdots,n_p)$.

In the case of $d$ dimensional Fermi liquid, the propoagator\cite{BG95}
\[
\bar g(x)=\frac{1}{(2\pi)^{d+1}}\int \frac{e^{-(k_0^2+{\mathcal E}({\bf k})^2)/{p_0}^2}e^{-\sqrt{-1}k x}}{-\sqrt{-1}k_0+{\mathcal E}({\bf k})} d^{d+1}k
\]
where ${\mathcal E}({\bf k})=({\bf k}^2-{p_F}^2)/2m$.

In the case of spatial 1$D$
\[
\bar g(x-y)=\sum_{\omega=\pm 1}\int\frac{e^{-\sqrt{-1}k(x-y)}} {-\sqrt{-1}k_0+\omega\cdot k}e^{-\sqrt{-1}p_F \omega\cdot(x-y)} d^2 k
\]
One performs multiscale decomposition of the scalar field
\[
\psi_x=\sum_{h=-\infty}^0 \psi_x^{(h)}
\]
where $\psi_x^{(h)}$ are Gaussian fields with propagators 
\[
\delta_{h h'}{\tilde C}(2^h p_0(x-y))2^{(d-2+\gamma)h},
\]
and $\psi_x^{(h)}$ is the same as that of $\psi_x^{(0)}$ suitably rescaled
\[
\psi_x^{(h)}=2^{(d-2+\gamma)h/2 }\psi_{2^h x}^{(0)}.
\]
Here the potential between fermions are assumed to be long range $J(x)\propto |x|^{-\alpha}$ and $\gamma=d+2-\alpha$.

The propagator for the Bose gas is
\[
g^{(t)}(x)=\frac{1}{(2\pi)^{d+1}}\int dk e^{\sqrt{-1}k x}\frac{t(k)}{-\sqrt{-1}k_0+\frac{{\bf k}^2}{2m}}
\]
where $t(k)$ is a positive cutoff function.

Consider the fields
\[
\chi_x^\pm=\frac{1}{\sqrt{2\rho}}(\psi_x^+\pm \psi_x^-)=\frac{1}{(2\pi)^{d-1}}\int dk e^{\pm\sqrt{-1}k x}\chi_k^{\pm}
\]
whose propagator has the form
\[
\langle \chi_x^\sigma \chi_y^{\sigma'}\rangle =\frac{1}{(2\pi)^{d+1}}\int e^{-\sqrt{-1}k x}t(k)G^{-1}(k)_{\sigma\sigma'},
\]
where the propagator matrix $G$ is
\[
G=\rho\left(\begin{array}{cc}
\frac{{\bf k}^2}{2m}&\sqrt{-1}k_0\\
-\sqrt{-1}k_0&-\frac{{\bf k}^2}{2m}\end{array}\right).
\]
Modification of the Gaussian measure due to the truncation yields\cite{BG95} $G'=G+2\rho\Delta t(k)$,
where 
\[
\Delta=\left(\begin{array}{cc}
2a& \sqrt{-1}c k_0\\
\sqrt{-1}c k_0&-2(b_0k_0^2+b{\bf k}^2)\end{array}\right),
\]
and
\[
G'=\rho\left(\begin{array}{cc}
\frac{{\bf k}^2}{2m}+4 a t(k)& \sqrt{-1}k_0[1+2 c t(k)]\\
-\sqrt{-1}k_0[1+2ct(k)]& -\frac{{\bf k}^2}{2m}-4(b k_0^2+b{\bf k}^2)t(k)\end{array}\right).
\]

The scale dependent $\Delta_{(h)}$ defined by
\[
\Delta_{(h)}=\rho\left(\begin{array}{cc} 2z_h\frac{p_0^2}{2m} & \sqrt{-1}d_h k_0\\
-\sqrt{-1}d_h k_0&-2\frac{2m}{p_0^2}k_0^2\zeta_h-2\alpha_h\frac{{\bf k}^2}{2m}\end{array}\right)
\]
yields
\[
G_{h'}(k)=\rho\left(\begin{array}{cc}
\frac{{\bf k}^2}{2m}+\frac{2p_0^2}{m}Z_{h'}&\sqrt{-1}k_0 E_h'\\
-\sqrt{-1}k_0 E_{h'}&-(\frac{{\bf k}^2}{2m}+\frac{8B_{h'}m k_0^2}{p_0^2}+\frac{2A_{h'}{\bf k}^2}{m})
\end{array}\right).
\]

The integral
\[
I=\int e^{-V(h')(\chi^{(\leq h'-1)} +\chi^{(h')})} P_{h'}(\chi^{(h')})\tilde P_{\leq h'-1}(d\chi^{(\leq h'-1)})
\]
agrees with that $h'$ interchanged by $h$, when 
the renormalization factor $Z_h(k)$, two beta function's coeficients $B_h(k)$, $A_h(k)$, effective potential $E_h(k)$, satisfy recursion relations
\begin{eqnarray}
Z_h(k)&=&Z_{h+1}(k)+t_h(k)z_h\nonumber\\
B_h(k)&=&B_{h+1}(k)+t_h(k)\zeta_h\nonumber\\
A_h(k)&=&A_{h+1}(k)+t_h(k)\alpha_h\nonumber\\
E_h(k)&=&E_{h+1}(k)+2 t_h(k) d_h\nonumber
\end{eqnarray}
with $Z_0=\epsilon t_0(k)$, $E_0=1$, $A_0=B_0=0$.

The square of the sound speed $c_h$ on scale $\gamma^h p_0$ is given by the ratio of the coefficients of ${\bf k}^2$ and $k_0^2$ in the scale $h$ propagator singularity, as follows\cite{BG95}
\begin{eqnarray}
c_h^2&=&\frac{4\frac{p_0^2}{2m} \frac{1}{2m} Z_h(1+4 A_h)}{E_h^2+16 B_h Z_h}\nonumber\\
&=&v_0^2Z_h\frac{1+4 A_h}{E_h^2+16 B_h Z_h}.\nonumber
\end{eqnarray}

At the lowest scale 
\[
c_0^2=\epsilon v_0^2=\epsilon (\frac{p_0}{m})^2, \quad \epsilon=Z_0=\lambda \hat v({\bf 0})\rho 2 m p_0^{-2}.
\]

We replace the potential 
\begin{eqnarray}
V(\psi)&=&\lambda\int_\Lambda v({\bf x}-{\bf y})\delta(x^0 -y^0)\psi_x^+ \psi_x^-
\psi_y^+\psi_y^- dx dy\nonumber\\
&&+\nu\int_\Lambda \psi_x^+\psi_x^- dx,\nonumber
\end{eqnarray}
where $\Lambda=[-\frac{1}{2}\beta,\frac{1}{2}\beta]\times[-\frac{1}{2}L,\frac{1}{2}L]^2$ to
\begin{eqnarray}
&&V(L^{(h)})=B_h^1 \sum_{L=L1,\cdots,L28}\chi(L^{(h)}) +B_h^2\sum_{L=L3,\cdots,L17 }\chi(L^{(h)})\nonumber\\
&&+A_h^1\sum_{L=L1,\cdots,L28} d{\bf S} (L^{(h)})+A_h^2\sum_{L=L3,\cdots,L17}
d{\bf S} (L^{(h)}),\nonumber
\end{eqnarray}
and $\Lambda=[0,\beta]\times[0,1]^2$.

In order to fix infrared cutoff on scale $\gamma^{-R}$ and ultraviolet cutoff scale 
$\gamma^{-U}$, we define Grassman field $\psi_x^\sigma$ and additional external field $\varphi_x^\sigma$ as\cite{BG95}.

For $k=(k_0,{\bf k})\in {\bf R}^3$, $e^{-\sqrt{-1}k_0\beta}=-1$, \quad $e^{-\sqrt{-1}{\bf k}\cdot{\bf L}}=+1$, and
\begin{eqnarray}
\psi_x^\sigma&=&\sum_k\frac{e^{\sqrt{-1}\sigma k x}}{\sqrt{\beta L}}
\frac{(e^{-k^2 \gamma^{-2U}}-e^{-k^2\gamma^{-2R}})^{1/2}}
{\sqrt{-\sqrt{-1} } k_0+{\mathcal E}({\bf k})}
 {\mathcal A}_k^\sigma,\nonumber\\
\varphi_x^\sigma&=&\sum_k\frac{e^{\sqrt{-1}\sigma k x}}{\sqrt{\beta L}} \epsilon_k^\sigma,\nonumber
\end{eqnarray}
where $x=({\bf x},t), k^2=k_0^2+{\mathcal E}({\bf k})^2$.

One performs the change of coordinates $\psi=\psi^{(\leq 0)}+\varphi_0$, and
\begin{eqnarray}
&&e^{-V_{eff}(\sqrt{Z_0}\varphi_0)}=c\int P^{(0)}_{Z_0}(d\psi^{(\leq 0)}e^{-V^{(0)}(\sqrt{Z_0}\psi^{(\leq 0)}+\varphi)}\nonumber\\
&&=e^{-\frac{1}{2}(\varphi_0,Z_0\Gamma_0^{-1} p^2\varphi_0)} \nonumber\\
&&\times c\int P_{Z_0}^{(0)}(d\psi)e^{-V^{(0)}(\sqrt{Z_0}\psi)} e^{(\psi,Z_0\Gamma^{-1}
p^2\varphi_0)},\nonumber
\end{eqnarray}
where $c$ is a formal normalization.

Their renormaization group use the gaussian measure $P_{Z_0}^{(h)}=Z_0^{-1}\Gamma_h(p)p^{-2}$ in momentum space with
\[
\Gamma_h(p)=e^{-(2^{-h}p/p_0)^2}.
\]
The parameter $p_0^{-1}$ can be interpreted as a lattice spacing in the infrared. In the ultraviolet $p_0$ was thought of as a physical mass. It should be fixed from experimental sound velocity or requirement of stability.

\begin{figure}
\begin{center}
\includegraphics[width=5cm,angle=0,clip]{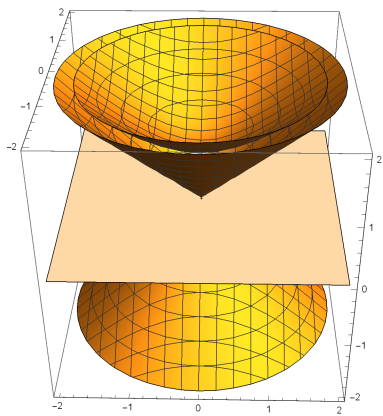} 
\end{center}
\caption{ The definition area of point form massive point-form wave functions. }
\label{lightcone}
\end{figure}

 The effective mass of Weyl fermion defines the propagation area of a fermion with an effective mass inside the light cone as shown in Fig.\ref{lightcone}.

Whether there is the phase transition of TR preserving and spin rotation (SR) symmetry breaking phase exists can in principle checked by temperature dependence of the thermal cycle\cite{CJR79}.  The detailed balance condition reads
\[
\frac{P(C,C')}{P(C',C)}=exp\{-\beta[S(C')-S(C)]\}
\]
We consider the difference of $C$ and $C'$ exists from the lattice nearest neighbour interactions, and
\[
P=P^{(i_N j_N)}\cdots P^{(i_2,j_2)}P^{(i_1,j_1)}
\]
If $\Delta S=S(C'(u'))- S(C(u))\geq 0$ a random number $r$ with uniform distribution between 0 and 1 
is generated and if $r<exp(-\beta \Delta S)$, the path running $u'$ will be 
adopted. In the ${\bf Z}_n$ model ($n=2,3,4,5,6,8$)\cite{CJR79}, hysteresis effects were observed.

\section{Summary and perspective}

Topologically non-trivial fields in gauge theory which are called instantons are proposed by `t Hooft\cite{tHooft76}, quantum fluctuations around multi-instanton fields were studied in \cite{BL79}, and reviewed by Polyakov\cite{Polyakov87}.

The idea was applied to spin models by Blatter et al.\cite{BBHN96}.  
The topological charge in $(2+1)D$,
$
Q=\frac{1}{8\pi}\int d^2 x\epsilon_{\mu\nu}{\bf S}\cdot(\partial_\mu{\bf S}\times \partial_\nu{\bf S})
$
was found to be a constant when the size of instanton is larger than $0.7a$.

Luescher\cite{Luescher82} discussed effects of instanton background in QCD.
Luescher's domain decomposition method for lattices with boundary matches Clifford algebra using real quaternions, since one can take Clifford pairs on the boundary with Gaussian average zero, and asymptotically $\beta_1'/\sqrt{-1}=0$ distributions. 

In order to perform lattice simulations of phonon propagations in fine lattices, renormalization group approach in momentum space following Benefatto and Gallavotti \cite{BG95} extended to the Clifford algebra would be appropriate. 

The recursive calculation of $Z_h, A_h,$ $B_h,E_h,$ $\mu_h,\nu_h$ in renormalization groups, starting from a $4\times 4$ lattice surrounded by Clifford pair bundaries to $2^{11}\times 2^{11}$ lattice surrounded by Clifford pair boundaries, using supercomputers is under investigation. 

 In $(2+1)D$ acoustics, existence of TR symmetry preserved and Spin Rotation symmetry broken phase is consistent with the APS index $n_+-n_-=2$.

Real quaternion lattice simulation of $(4+1)D$ and to $(6+1)D$ systems and extending spin systems to gauge systems remain as future studies.  

\begin{acknowledgements}
I thank Dr. Serge DosSantos at INSA for valuable informations on NDT and Prof. M. Arai for supports. Thanks are also due to the RCNP of Osaka University for allowing use of super computers there, and Tokyo Institute of Technology for consulting references.
\end{acknowledgements}


\end{document}